\documentclass{aa}  

\usepackage{graphicx}
\usepackage{txfonts}
\usepackage[]{hyperref}

\newcommand{\teff}{T$_{\rm eff}$}
\newcommand{\Msun}{M$_{\rm \odot}$}

\usepackage{xcolor}
\hypersetup{
    colorlinks,
    linkcolor={red!50!black},
    citecolor={blue!50!black},
    urlcolor={red!80!black}
}
\begin{document}

   \title{Mass loss along the red giant branch of the intermediate stellar populations in NGC\,6752 and NGC\,2808}

   \subtitle{}

   \author{M. Tailo
          \inst{1,2}
          \and
          A.\,P.\,Milone
          \inst{1,3}
          \and
          A.\,F.\,Marino
          \inst{1}
          \and
          F.\,D'Antona
          \inst{2}
          \and
          M.\,V.\,Legnardi
          \inst{3}
          \and
          T.\,Ziliotto
          \inst{3}
          \and
          E.\,P.\,Lagioia
          \inst{4}
          \and
          S.\,Jang
          \inst{5}
          \and
          E.\,Bortolan
          \inst{3}
          \and
          P.\,Ventura
          \inst{2}
          \and
          C.\,Ventura
          \inst{2}
          \and
          E.\,Dondoglio
          \inst{1,3}
          \and
          F. Muratore
          \inst{3}
          \and
          A. Mohandasan
          \inst{3}
          \and
          M.\,Barbieri
          \inst{3}
          \and
          S.\,Lionetto
          \inst{3}
          \and
          G.\,Cordoni
          \inst{6}
          \and
          F.\,Dell'Agli
          \inst{2}
          }

   \institute{
        Istituto Nazionale di Astrofisica- Osservatorio Astronomico di Padova, Vicolo dell’Osservatorio 5, Padova, IT-35122\\
        \email{mrctailo@gmail.com, marco.tailo@inaf.it}        \and
        INAF, Observatory of Rome, Via Frascati 33, 00077 Monte Porzio Catone (RM), Italy 
        \and
        Dipartimento di Fisica e Astronomia “Galileo Galilei”, Univ. di Padova, Vicolo dell’Osservatorio 3, Padova, IT-35122
        \and
        South-Western Institute for Astronomy Research Yunnan University, Kunming, 650500, P.R. China
        \and
        Center for Galaxy Evolution Research and Department of Astronomy, Yonsei University, Seoul 03722, Korea        
        \and
        Research School of Astronomy and Astrophysics, Australian National University, Canberra, ACT 2611, Australia
        }

   \date{Received September XXXX; accepted March YYYY}

 
  \abstract{The morphology of the Horizontal Branch (HB) in Globular Clusters (GC) is among the early evidences that they contain multiple populations of stars. Indeed, the location of each star along the HB depends both on its initial helium content (Y) and on the global average mass loss along the red giant branch ($\mu$). In most GCs, it is generally straightforward to analyse the first stellar population (standard Y), and the most extreme one (largest Y), while it is more tricky to look at the "intermediate" populations (mildly enhanced Y).
  In this work, we do this for the GCs NGC\,6752 and NGC\,2808; wherever possible the helium abundance for each stellar populations is constrained by using independent measurements present in the literature. 
  We compare population synthesis models with photometric catalogues from the \textit{Hubble Space Telescope} Treasury survey to derive the parameters of these HB stars. 
  We find that the location of helium enriched stars on the HB is reproduced only by adopting a higher value of $\mu$ with respect to the first generation stars in all the analysed stellar populations. We also find that $\mu$ correlates with the helium enhancement of the populations. This holds for both clusters. This finding is naturally predicted by the model of ``pre-main sequence disc early loss'', previously suggested in the literature, and is consistent with the findings of multiple-populations formation models that foresee the formation of second generation stars in a cooling flow.}


   \maketitle
%

\section{Introduction}
\label{sec:intro}
Already in the sixties of the 20th century, it was clear that the horizontal branch (HB) stars in Globular Clusters (GC) are the descendant of red giants, and the HB is a locus of stars differing in mass. The violent He–ignition in the degenerate He–core  (He–flash, see e.g. \citealt{sh1962} and \citealt{hs1964}) gives origin to a double–source structures (central He–burning plus H–burning shell). The appearance of an “average” HB can be easily reproduced by assuming an appropriate range of mass loss along the red giant branch (RGB) phase, therefore assuming that the total mass of HB stars as a free parameter \citep{rood1970, ir1970}. At the same time, it was recognized that HB colour distribution does not depend only on metal content, as expected by \cite{faulkner1966}  but that a "second parameter" was required \citep{vdb1967,sw1967}. The reasons for a mass spread in this phase were immediately connected to the spread in the mass lost by the stars evolving along the red giant branch and suffering the helium flash before settling in the stage of burning helium in the core and hydrogen in the shell on the “zero age horizontal branch” (ZAHB).
The hunt for a second and a third parameter \citep[e.g.][]{fusipecci1993,fpb1997,dotter_2011,milone_2014}
provided interesting hints, but not completely satisfactory results.

The discovery of extremely blue HB (bHB) members added more difficulties, until the observational situation suffered a dramatic change, following the discovery (sometimes re--discovery) of complex abundance distributions and completely new photometric features in CM diagrams (multiple Main Sequences (MS), multiple subgiant and giant branches).
Indeed, almost all GCs observed up to day exhibit, in a large fraction of its population, the effects of proton burning at high temperature, effects manifested by correlations and anti--correlations among light elements (Na--O, Mg--Al, etc) - we quote for the recent literature  \citet{milone_2017,bastian_2018,Gratton_2019,marino_2019} and \citet{Milone_2022} and references therein - and, in a few cases, by enhanced s--process elements - see \citet{marino2009, villanova2010, carretta2011, marino2012m22}.
These observations conclusively showed that GCs are not simple  population systems, but host at least a couple of star  generations, which hereinafter we will name ``first generation" (1G) and ``second generation" (2G).

The gas subject to p--captures forming the second generation stars is expected to have also an increased helium mass fraction (Y), which also affects the HB morphology. Along the HB, stars of smaller mass populate hotter \teff's locations: both an increase in mass loss and Y work along the same direction by decreasing the evolving mass, so a spread in Y reduces the mass loss spread needed to fit an extended HB \citep{dantona_2002}.  
Therefore, considering the presence of at least two stellar populations has been mandatory to provide a simple key to understand some of the puzzles provided by the distribution of stars along the branch, e.g. the non monotonic distribution of stars. The prime example of this is NGC\,2808, which is characterized by a distinctly bimodal distribution in B--V colors (including a relatively small number of RR Lyrae variables, compared to the bHB and red HB (rHB) populations, see e.g. \citealt[][]{catelan1998}). In such cases, a unimodal mass distribution, even requiring a very large mass spread of about 0.1\Msun, is not compatible with the distribution, while different helium content of the different stellar groups populating the cluster may be more appropriate \citep{dantona_2005}.

The study of HB then implies dealing with a large number of parameters at the same time. But, while metallicity\footnote{Usually expressed as the iron content, [Fe/H], together with the alpha elements ratio, [$\alpha$/Fe].} and cluster age can be evaluated independently (see \citealt{dotter_2010,dotter_2011,VandenBerg_2013,marinf_2009} for examples of age determination and \citealt{carretta_2009,Carretta2010,carretta_2015} for examples of metallicity determination), in the traditional approach helium and RGB mass loss are derived by directly studying the HB morphology itself. This implies a parameter degeneracy that can be challenging to resolve. 

In a series of papers, thanks to the capabilities of the Wide Field Camera 3 (WFC3) on board the \textit{Hubble Space Telescope} (HST), \cite{milone_2013,milone_2015,milone_2018} provided estimates of the helium abundances of the various stellar populations hosted in a large number of GCs, allowing the community to break the parameter degeneracy and making a full description of HB stars finally possible. On the basis of these observational studies, \citet[][hereinafter T20]{tailo_2020} and \citet[][hereinafter T21]{tailo_2021} examined a sample of 56 GCs showing the role of each parameter in shaping the HB morphology. 

One important, although intriguing, result in T20 was the necessity to increase the mass loss of the most extreme part of the second generation stars (2G and 2Ge, respectively), with respect to the first generation one (1G), in order to correctly describe their position along the HB in almost all clusters of the sample\footnote{The nomenclature, here and in the following, follows T20, T21 and the other works from Milone and collaborators. 1G stands for `first generation', i.e. the population of stars that has chemical pattern compatible with the field stars of the same metallicity. In this context, 2G (second generation) collectively indicates the group of stellar populations with signatures of hot proton capture nucleosynthesis. Finally, 2Ge indicates the part of the 2G with the most altered chemistry in each cluster.}. Possible differences in mass loss among stars belonging to the different populations was previously suggested, for selected GCs, by many authors (for a partial list see e.g. \citealt{dantona_2002, dantona_2008, Salaris2008ApJ...678L..25S, dalessandro_2011, dalessandro_2013, Cassisi2014A&A...571A..81C}), sometimes with conflicting results. The large sample analysed by Tailo and collaborators showed that the \textit{difference} in mass loss between the 2Ge and the 1G is clearly correlated with both the present-day \citep{baumgardt_2018} and initial mass \citep{baumgardt_2019} of the host clusters, with the helium abundance of the populations and, generally, with the parameters connected to the complexity of the MPs phenomenon. On this basis, Tailo and collaborators suggested that these correlations are the fossil traces of the MPs formation mechanism.

Granted the question is still debated and no conclusive description is available, according to the most successful scenarios, the formation of the 2G happens after the gas in the cluster has been polluted with the product of high temperature proton-burning \citep[a list of wide scope descriptions of the various mechanism and scenarios can be found in e.g.][]{renzini_2015,bastian_2018,Gratton_2019,Milone_2022}. In the context of a cooling flow scenario \citep{dercole_2008}, the 3D hydrodynamical simulations by
\cite{calura_2019} show that the most helium enhanced population forms in a denser environment with respect to the 1G. The different environment possibly allows to attain the needed additional mass loss, via a process of early magnetic disc destruction \citep[in the T-Tauri stage, see][and references therein]{Armitage1996MNRAS.280..458A, Bouvier1997A&A...326.1023B, tailo_2015, tailo_2020}. More massive clusters form their 2Ge in an even denser environment, causing, on average, an even earlier loss of the pre-main sequence disc, thus providing higher rotational velocities and consequentially larger differences in mass loss. 

One of the missing evidences needed to build any self-consistent theoretical scenario, is the behaviour of the intermediate stellar populations, i.e. those 2G populations less enhanced than the 2Ge. However, independent evaluations of helium abundance of the intermediate 2G stars, needed to avoid the degeneracy problem, have only been conducted in a few clusters. 
\citet{milone_2013} and \cite{milone_2015} have provided the estimates for the helium abundance of the intermediate populations, respectively for NGC\,6752 and NGC\,2808, giving us the chance to describe the entire HB of these two clusters. For the cluster NGC\,2808, in particular, we considered in T20 only the 1G and the extreme tail of the HB, the so called "blue hook" stars \citep[bhk]{moehler2004}, so we here examine the whole 2Ge population, consistent of two separate HB clumps named EBT2 and EBT3 \citep{bedin2000}. 
The aim of this complementary work is then to describe the location of these stellar populations on the HB and analyse their behaviour to add information that will able to self-consistently confirm the proposed scenario of pre--main sequence disk early loss. 

The present work is divided into four main parts. In \S\,\ref{sec:data_and_models}, we present the data and the models we employ, as well as the helium abundance estimates we use. In  \S\,\ref{sec:6752} and \S\,\ref{sec:2808}, we describe our results concerning the HB stellar populations in both NGC\,6752 and NGC\,2808. Finally, in \S\ref{sec:disc} and \S\,\ref{sec:concl}, we discuss the combined results for both clusters in the context of the MPs formation scenarios and present a summary of our findings.

\section{Data and models}
\label{sec:data_and_models}

To estimate the RGB mass loss of the intermediate stellar populations in NGC\,6752 and NGC\,2808, we combine multi-band photometry from the HST, helium abundances derived from the study of the photometric data and suitable stellar models built for GC stars with enhanced helium content. The following subsections (\ref{sub:data} to \ref{sub:models}) describe the photometry, the helium abundances and the theoretical models.

\subsection{Data}
\label{sub:data}

We exploit the photometric and astrometric catalogs from the \textit{HST} UV legacy survey of GCs \citep{piotto_2015, nardiello_2018}. These catalogues include accurate astrometry and photometry in the F275W, F336W, and F438W bands of the ultraviolet and visual (UVIS) channel of the Wide Field Camera 3 (WFC3) and in the F606W and F814W bands of the Wide Field Channel of the Advanced Camera for Surveys (ACS) on board the \textit{HST}. We refer to \citet{piotto_2015} and \citet{nardiello_2018} for basic details on the data set and the data reduction procedure. The photometric catalogue of NGC\,2808 has been corrected for differential reddening following the recipe described in \cite{milone_2012c} while for NGC 6752, which is poorly affected by differential reddening, we used the original photometry. 

\subsection{Helium abundance and population ratios}
\label{sub:elio}

We refer to \cite{milone_2013} and \citet[][]{milone_2015} to get the helium abundance estimates of the intermediate populations in both NGC\,6752 and NGC\,2808 and be able to break the degeneracy associated to the HB stars in these clusters.

\begin{table}[]
    \centering
    \caption{The individual stellar populations in NGC\,6752 and NGC\,2808}
    \begin{tabular}{ccc}
        \hline
        \hline
         ID &$\delta Y$& $\rm N/N_{tot}$ \\
        \hline
        \multicolumn{3}{c}{NGC\,6752}\\
        \multicolumn{3}{c}{Age=13.00, [Fe/H]=-1.54}\\
        \hline
         Red     & 0.00  &$0.31\pm0.03$\\
         Middle  & 0.008 &$0.41\pm0.02$\\
         Blue    & 0.026 &$0.28\pm0.03$\\
         \hline
        \multicolumn{3}{c}{NGC\,2808}\\
        \multicolumn{3}{c}{Age=12.00, [Fe/H]= -1.14}\\
         \hline
         A & $-0.033\pm0.003$*& $0.058\pm0.005$ \\       
         B & $0.000\pm0.002$  & $0.174\pm0.009$ \\        
         C & $0.002\pm0.004$  & $0.264\pm0.012$ \\        
         D & $0.040\pm0.005$  & $0.313\pm0.013$ \\        
         E & $0.09\pm0.005$   & $0.191\pm0.010$ \\        
        \hline
    \end{tabular}
    \tablefoot{Columns are: name of the population or sequence (ID), helium enhancement ($\delta$Y), number fraction ($\rm N/N_{tot}$). See \citet{milone_2013},\citet{harris_1996,harris_2010},\citet{Yong2005},\citet{Carretta2010}, T20,\citealt{marino_2014},\citet{milone_2015},\citealt{carretta_2015}. *see text for details}
    \label{tab:pop_individual}
\end{table}

\subsubsection{NGC6752}
The cluster NGC\,6752 ([Fe/H]=-1.54 as per \citealt{harris_1996,harris_2010} catalogue, \citealt{Yong2005} and \citealt{Carretta2010} and age of $13.0\pm0.50$\,Gyr, as per T20) was studied photometrically by \cite{milone_2013}, who found that it hosts three main populations, divided into distinct sequences that can be followed from the low main sequence up to the tip of the RGB. By combining their photometric analysis with the spectroscopic results from \cite{Yong2003A&A...402..985Y,Yong2008ApJ...684.1159Y}, Milone and collaborators found that NGC\,6752 hosts three stellar populations, in the number proportion listed in Table\,\ref{tab:pop_individual}. For the two 2G sequences, Milone and collaborators found the values of helium mass fraction enhancement ($\rm \delta Y$), with respect to the 1G population, we list in \ref{tab:pop_individual}.Further details on the analysis can be found in \cite{milone_2013}.
The population ratios are also compatible with the percentage of $0.294\pm0.023$ found for the 1G in \cite{milone_2017}. 

\subsubsection{NGC2808}
The stellar populations in NGC\,2808 ([Fe/H]$\sim$-1.14, see \citealt{harris_1996,harris_2010} catalogue, \citealt{Carretta2010}, \citealt{marino_2014}, \citealt{carretta_2015} and age of $12.0\pm0.75\,$Gyr, as per T20) are more complex. Indeed, \cite{milone_2015} described five main stellar populations in this cluster, dubbed ABCDE, observed in the main sequence, up to the Tip of the RGB. \cite{milone_2015} also performed a spectro-photometric analysis of these five populations, finding that they have enhanced helium and altered chemistry. 
We report the percentages and the values of $\rm \delta Y$ with respect to group B in the lower part of Table \ref{tab:pop_individual}. 
The combined analysis of the ChM by \cite{milone_2017} indicates that Populations A and B are the total of the 1G stars, representing $0.232\pm0.014$ of the total stars in the cluster. 

Alternatively, given also the results from \citet[][and references therein]{marino_2019,Marino2019ApJ...887...91M,Legnardi_2022}, group A can be interpreted as having the same helium abundance and light elements ratio as group B but a slightly different value of [Fe/H], higher for group A. As it will be clear in \S\,\ref{sub:2808_rhb}, we plan to disentangle the 1G from the 2G on the rHB in this cluster by means of the spectroscopic observation by \cite{marino_2014}. With these observations we cannot have an amount of stars belonging to Population A large enough to have a satisfying fit in isolation, therefore we forgo the distinction between the two parts of the 1G and study them as a single stellar population. 

\subsection{Models}
\label{sub:models}

We adopted the stellar-evolution models and the isochrones from T20 and T21. The models themselves are obtained with the \textsc{aton} 2.0 stellar evolution code by \cite{ventura_1998} and \cite{mazzitelli_1999}. The models span a large range of [Fe/H], [$\rm \alpha/Fe$] and Y values to accommodate the variations observed among the Galactic GC population. The individual HB tracks are followed until they reach $\rm He_c<=0.01$, thus right before the proper end of the core helium burning. 

We compare the photometric data of NGC\,6752 and NGC\,2808 with grids of synthetic CMDs derived from the appropriate models. The individual simulations in our grids are calculated following the recipes of \citet[][and references therein]{dantona_2005} and includes a number of HB stars (2000) large enough to avoid any variance problem. The mass of each HB star ($\rm M^{HB}$) is chosen as follows: $\rm M^{HB}=M^{Tip}(Z, Y, A) - \Delta M(\mu,\delta)$; where $\rm M^{Tip}$ is the stellar mass at the RGB tip, function of age (A), metallicity (Z) and helium content (Y), $\rm \Delta M$ is the mass lost by the star, described by a Gaussian profile with central value $\mu$ and standard deviation $\delta$, which will be our sought of average  integrate mass loss and its spread\footnote{We use the Gaussian profile to the mass loss for it's ease of use and consistency with the other paper works from our group, however other kind of shapes for the mass loss can work. See e.g. \cite{Cassisi2014A&A...571A..81C} for a case where the modelling is done with a flat distribution.}. Finally, we simulate the effect of metal levitation between the \cite{grundahl_1999} and \cite{momany2012a} jumps by using super solar bolometric correction when the stars reach appropriate temperatures, as suggested by the results of \cite{grundahl_1999,dalessandro_2011,brown_2016,tailo_2017}.

\section{The HB of NGC6752}
\label{sec:6752}

\begin{table}
    \centering
    \caption{Parameters of the HB simulation grids used for NGC\,6752. }
    \begin{tabular}{cccc}
        \hline
        \hline
        ID & Y &$\rm \mu_{range}/M_\odot$ &$\rm \delta_{range}/M_\odot$ \\
        \hline
        1G   & 0.250 & --           & --             \\  
        2G1  & 0.258 & 0.10 -- 0.30 & 0.002 -- 0.250 \\  
        2Ge  & 0.275 & 0.10 -- 0.30 & 0.002 -- 0.010 \\ 
        \hline
        \multicolumn{4}{c}{Alternate iHB description}\\
        \hline
        2G1a & 0.255 & 0.10 -- 0.30 & 0.002 -- 0.010 \\  
        2G1b & 0.270 & 0.10 -- 0.30 & 0.002 -- 0.010 \\  
        \hline
        \hline
    \end{tabular}
    \tablefoot{Columns are: The ID of the population and its value of Y, the range of $\rm \mu$ and $\rm \delta$ values adopted. For the $\rm \mu$ and $\rm \delta$ ranges we adopt a step of $\rm 0.003 M_\odot$ and $\rm 0.001 M_\odot$ respectively.}
    \label{tab:grids_6752}
\end{table}

\begin{table*}
    \centering
    \caption{Parameters of the HB population in NGC\,6752. }
    \begin{tabular}{ccccccccc}
        \hline
        \hline
        ID & Y &N & N/N$_{\rm tot}$& $\rm \mu/M_\odot$&$\rm \delta/M_\odot$&$\rm M_{Tip}/M_\odot$&$\rm \bar{M}_{HB}/M_\odot$&$\rm \Delta\mu/M_\odot$\\
        \hline
        1G   & 0.250    & $37\pm 5$   &$32.2\pm 4.2\%$    &$0.216\pm 0.022$   &$0.006\pm 0.001$   &0.814  &$0.598\pm 0.022$   & -- \\  
        2G1  & 0.258    & $53\pm 5$   &$46.1\pm 4.6\%$    &$0.246\pm 0.028$   &$0.017\pm 0.003$   &0.801  &$0.555\pm 0.028$   & $0.03\pm0.015$\\  
        2Ge  & 0.275    & $27\pm 4$   &$23.5\pm 3.7\%$    &$0.285\pm 0.024$   &$0.004\pm 0.002$   &0.768  &$0.483\pm 0.024$   & $0.069\pm0.016$\\ 
        \hline
        \multicolumn{9}{c}{Alternate iHB description}\\
        \hline
        2G1a & 0.255    & $37\pm 5$   &$32.2\pm 4.3\%$   &$0.246\pm 0.020$   &$0.007\pm 0.002$   &0.808  &$0.562\pm 0.020$   & $0.03\pm0.013$\\  
        2G1b & 0.270    & $17\pm 6$   &$14.8\pm 3.1\%$   &$0.270\pm 0.021$   &$0.005\pm 0.002$   &0.787  &$0.517\pm 0.021$   & $0.054\pm0.015$ \\  
        \hline
        \hline
    \end{tabular}
    \tablefoot{Columns are: name of the population (ID), helium abundance (Y), number of stars and number fraction (N and $\rm N/N_{tot}$), average mass loss ($\mu/M_\odot$), mass loss spread ($\rm \delta/M_\odot$), mass of the star at the Tip of the RGB ($\rm M_{Tip}/M_\odot$), average HB mass for the star in the population ($\rm \bar{M}_{HB}/M_\odot$) and mass loss difference compared to the 1G stars ($\rm \Delta\mu/M_\odot$).}
    \label{tab:par_hb_6752}
\end{table*}

In this section we will describe the analysis of the HB in NGC\,6752, starting from its morphology (\S\,\ref{sub:morph_6752}) and then analysing its stellar populations. We start from the two extreme ones (the 1G and the 2Ge,\S\,\ref{sub:extreme_6752}) and then we will model the intermediate branch (\S\,\ref{sub:intermediate_6752}). 

\subsection{Morphology}
\label{sub:morph_6752}

\begin{figure*}
    \centering
    \includegraphics[width=2.0\columnwidth,trim={0cm 1cm 0cm 0cm}]{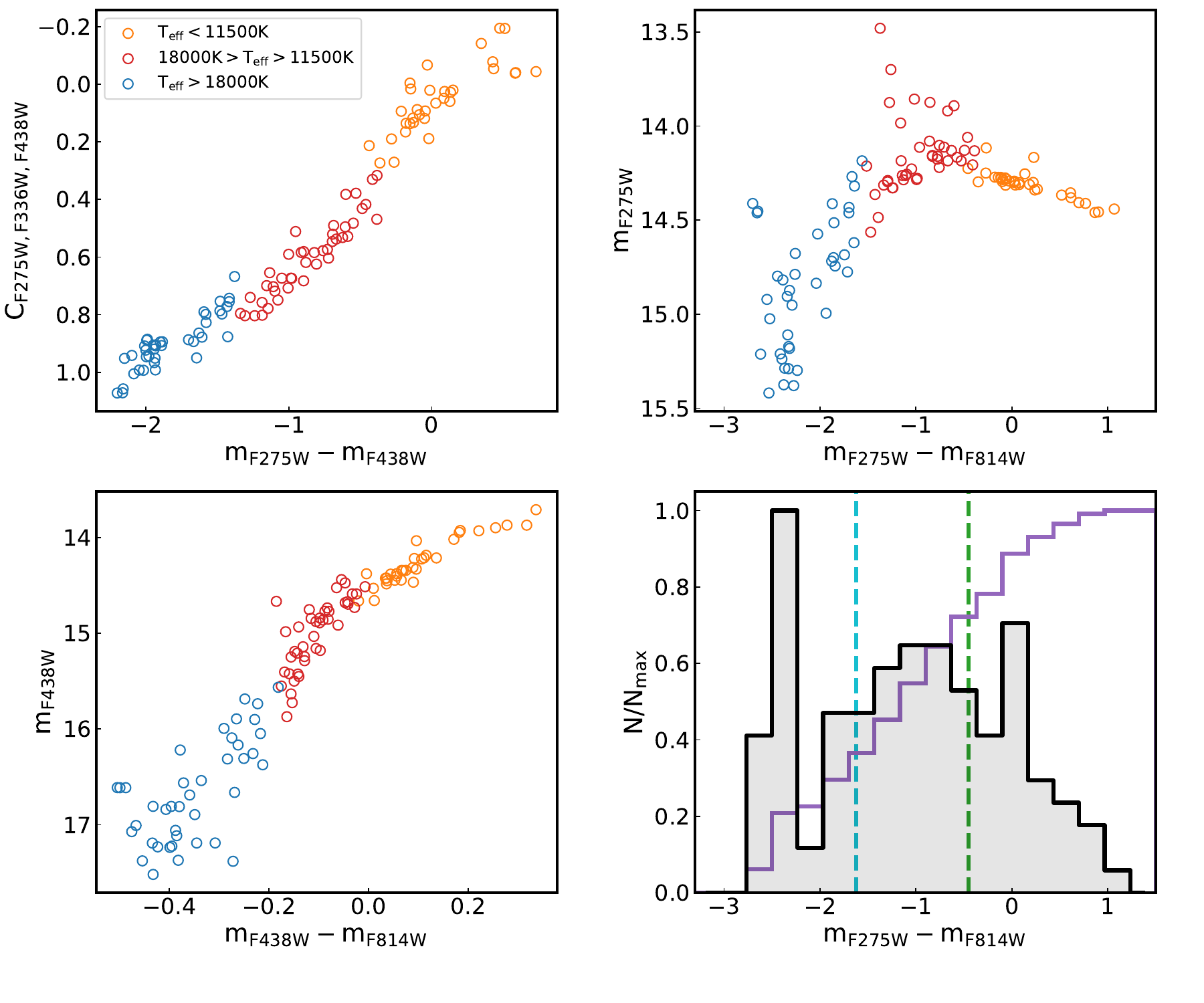}
    \caption{\textit{Top left panel:} $\rm C_{F275W,F336W,F814W}$ vs $\rm m_{F275W}-m_{F438W}$ pseudo two-color diagram of the HB stars in NGC\,6752. Stars with T<11500 K, T>18000 K, and 11500 K < T  < 18000 K are marked with orange, blue, and red points. \textit{Top right panel:} $\rm m_{F275W}$ vs. $\rm m_{F275W}-m_{F814W}$ CMD of the HB stars in NCG\, 6752. \textit{Bottom right panel:} Histogram of the colour distribution of the photometric data. The purple profile represents the cumulative distribution of the points. The two dashed lines, green and cyan, mark the position of the G- and M- jumps, respectively.\textit{Bottom left panel:} $\rm m_{F438W}$ vs. $\rm m_{F438W}-m_{F814W}$ CMD of the HB stars in NCG\, 6752. Where applicable, we plot stars in the three ranges of temperature with the same colour coding.} 
    \label{pic:morph_6752}
\end{figure*}

Our sample for this cluster consists of 115 HB stars which all populate the part of the branch bluer of the instability strip (IS). \cite{milone_2014} found a value of $\sim$0.38 mag for the parameter $\rm L_1$\footnote{The distance in colour between the RGB and the red part of the HB} and if we follow the separation introduced in T20, we can describe NGC\,6752 as a M13-like cluster\footnote{Those cluster with $\rm L_1$ higher than 0.45 mag, see Figure 5 in T20}. Being composed of blue stars only, this particular HB is best studied with combinations of optical and UV filters as we show in Figure \ref{pic:morph_6752}.

Two important discontinuities are observed along the HB of every GC and NGC\,6752 is no exception: the \citet{grundahl_1999} and \cite{momany_2004} jumps, hereinafter G- and M- jumps. These are universal features of the HB locus, and are connected to changes of the internal structure of the HB stars with temperature\footnote{More specifically, the changes of the two external convective layers and their interplay with radiative levitation, see \citet[][]{grundahl_1999,momany_2004,brown_2016,brown_2017,tailo_2017} and references therein.}. 
The G-jump and the M-jump are located at $\sim$11500 and $\sim$18000K \citep[][]{grundahl_1999,momany_2004,brown_2016,brown_2017}, respectively, and, as Brown and collaborators showed, they are most easily identified in the $\rm C_{F275W,F336W,F438W}$\footnote{$\rm C_{F275W,F336W,F438W}=m_{F275W}-m_{F336W}-(m_{F336W}-m_{F438W})$} vs.  $\rm m_{F275W}-m_{F438W}$ diagram. 

We show this diagram for NGC\,6752 in the upper left panel of Fig.\,\ref{pic:morph_6752}.
Even if the HB in NGC\,6752 is populated without apparent gaps and its metallicity value is low ([Fe/H]=-1.54), the transitions associated with the jumps can be appreciated by eye, in spite of them being not sharp. We colour in orange the stars cooler than the G-jump, in blue those hotter than the M-jump and in red the ones in between the two. Therefore, we identify $32$ stars before the 11500K mark and $39$ stars in the region past 18000K. As a consequence $41$ stars populate the region between the two jumps. In the following we use this separation to comment on the morphology of the HB of this cluster.

We report the $\rm m_{F275W}$ vs. $\rm m_{F275W}-m_{F814W}$ CMD in the upper right panel of Figure \ref{pic:morph_6752}. This is one of the best combinations of filters in the \textit{HST} Treasury survey to study bHB stars as it is the one with the largest sensitivity to temperature \citep{piotto_2015,milone_2015,milone_2017,milone_2018}. We also report the histogram of the colour distribution of the points (black) as well as their cumulative distribution (red) in the bottom right panel of Figure \ref{pic:morph_6752}. From the figure we see there that the distribution of the HB stars in NGC\,6752 has one narrow peak at $\rm m_{F275W}-m_{F814W}\sim -2.5$, a group of stars broadly distributed at $\rm m_{F275W}-m_{F814W}\sim -1$ and a final peak at $\rm m_{F275W}-m_{F814W}\sim 0 $. The cumulative distribution shows that the bluest peak hosts $\sim 20\%$ of stars, the reddest one $\sim 20\%$ as well, while the majority ($\sim 60\%$) of the HB stars are in the central group.
With the separation in temperature we have done in the upper left panel, we locate the position of the G-jump at $\rm m_{F275W}-m_{F814W} \sim -0.45$ and the M-jump at $\rm m_{F275W}-m_{F814W}\sim -1.62$ in the CMD. There seems to be no noticeable gaps at these locations in the HB. We also report the position of the jumps in the lower right panel of Figure \ref{pic:morph_6752} as the green and purple dashed lines, respectively for the G- and M-jump. For completeness, we also plot the $\rm m_{F438W}$ vs. $\rm m_{F438W}-m_{F814W}$ CMD in the lower right panel of Figure \ref{pic:morph_6752}. In this CMD the locations of the G and M jumps are $\rm m_{F438W}-m_{F814W}\sim -0.03$ and -0.19, respectively. 
All this procedure ensures us that all the peaks in the colour distribution are independent from the G- and M- jumps position, and allows us to associate the peaks with the three observed stellar populations.

\subsection{Mass loss of the first and the extreme second generation stars}
\label{sub:extreme_6752}

\begin{figure}
    \centering
    \includegraphics[width=\columnwidth]{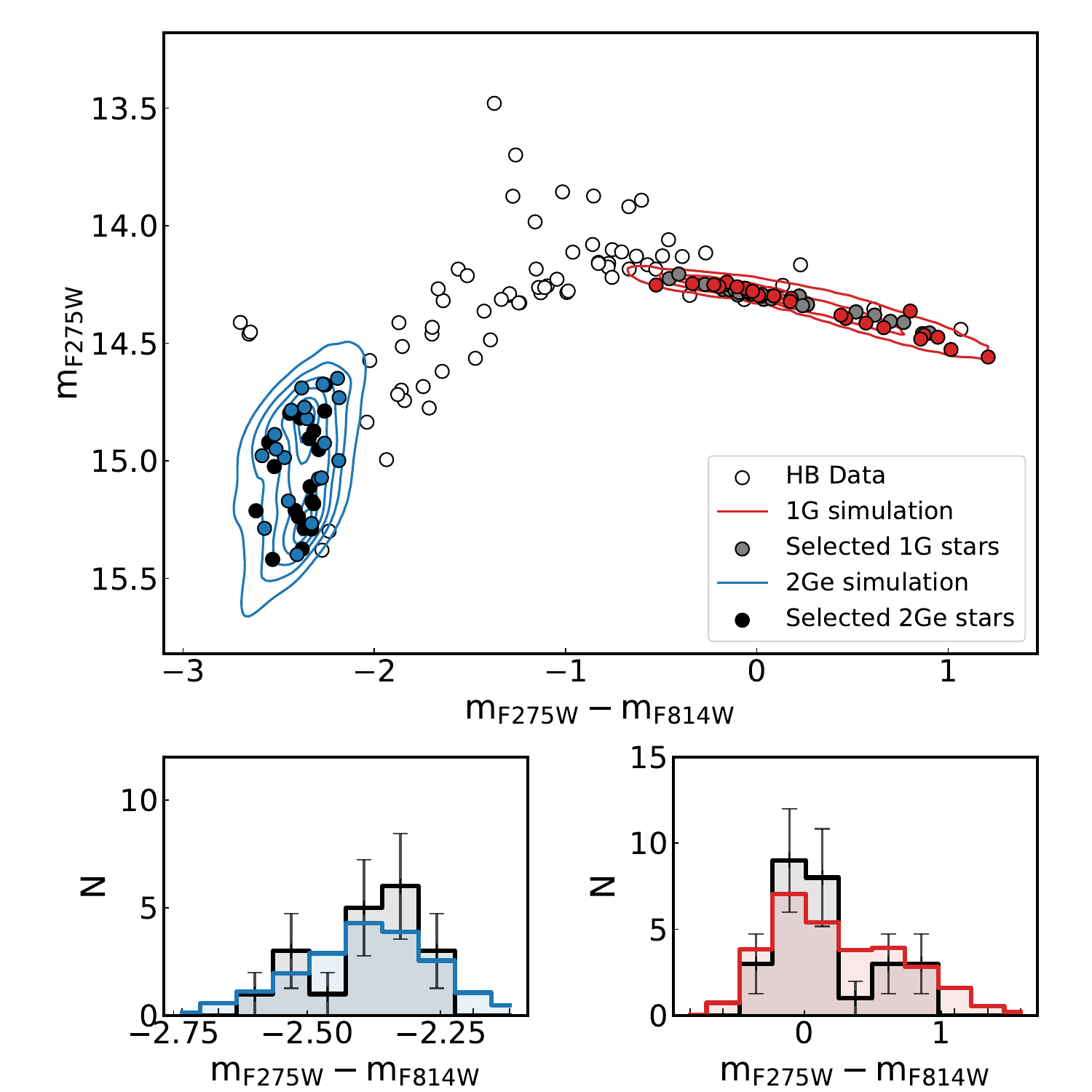}    
    \caption{Best fit simulations for the 1G and 2Ge stars in NGC\,6752. The grey and black points are the stars belonging to the 1G and the 2Ge, respectively, selected according to the simulation area (red and blue respectively for the 1G and the 2Ge). The coloured points, colour coded as the simulations, are a typical realisation of each population. Histograms in the bottom panels compare the colour distribution of the two simulations (red and blue) and the selected stars (black). The simulation histograms are normalized to the number of stars included in each population.}
    \label{pic:extremes_6752}
\end{figure}

\begin{figure}
    \centering
    \includegraphics[width=\columnwidth]{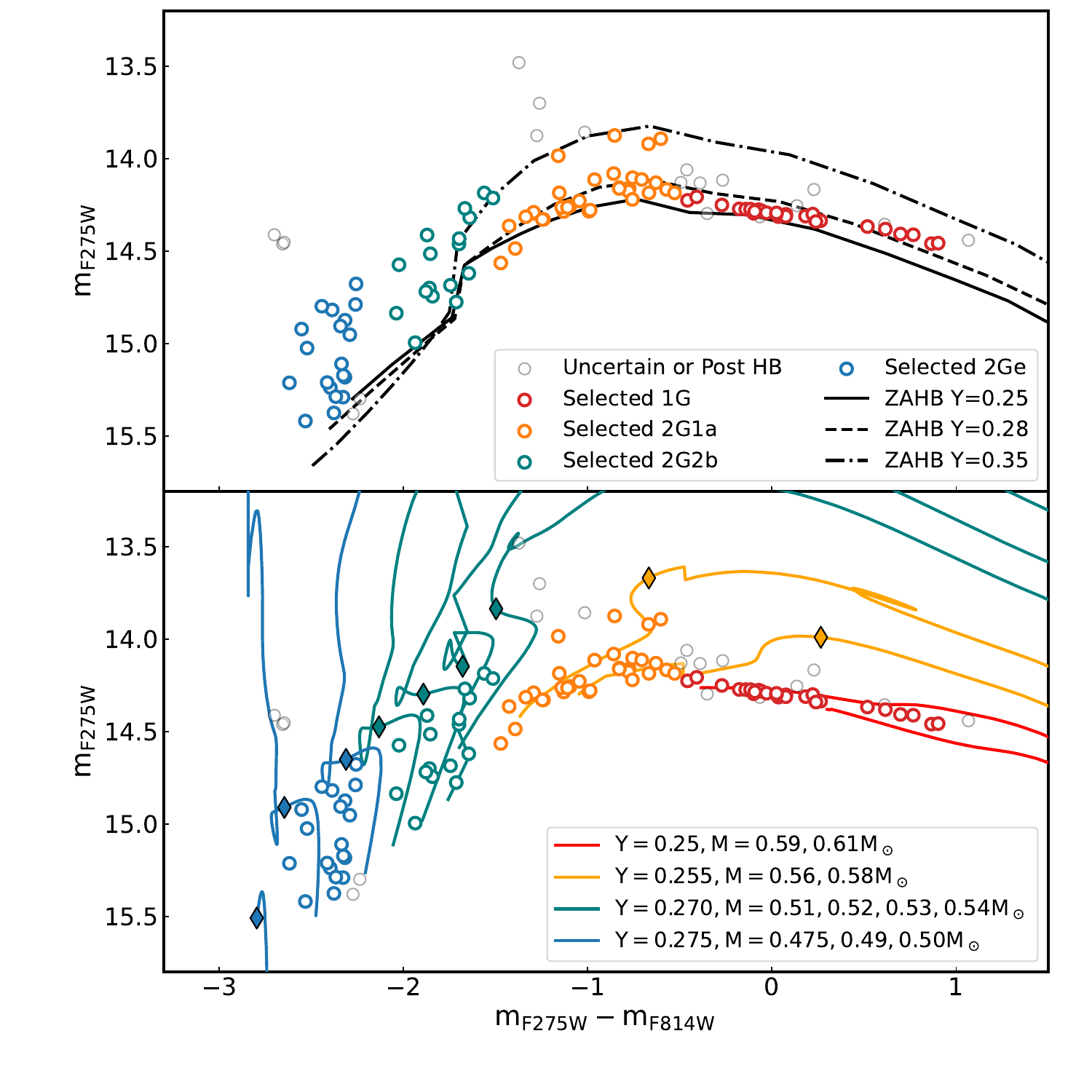}    
    \caption{\textit{Top panel:} $\rm m_{F275W}$ vs. $\rm m_{F275W}-m_{F814W}$ CMD of the HB in NGC\,6752. Each population is highlighted with the same colour code used in other part of this paper. The three lines represents the ZAHB loci of appropriate metallicity and Y=0.25,0.28,0.35, respectively for the solid, dashed, and dot-dashed one. \textit{Bottom panel:} As the top panel but we show the stellar tracks involved in the simulations to show their evolution during and past core helium burning. For each track, the end of the core helium burning is highlighted by the coloured, solid diamonds. Tracks are coloured as their respective stellar populations.}
    \label{pic:tracks_6752}
\end{figure}

We refer to the results from T20 for what concerns the 1G on the HB of NGC\,6752; however, since the assumptions we make here for the 2Ge are slightly different from the ones we made in our past work, to ensure consistency, we will study these stars again. For the 1G, its colour distribution is best-fit by a mean mass loss value of $\rm \mu= 0.216\pm 0.022 M_\odot$ and mass-loss spread value of $\rm \delta= 0.006\pm 0.001 M_\odot$. We use these value to identify and study the 1G stars on the branch.

The 1G simulation corresponding to the selected parameters is reported as the red contour plot in the upper panel of Figure \ref{pic:extremes_6752}, while the solid, red points represent a typical realisation of the simulation. To find which stars belong to the 1G, we use the best fit simulation and calculate its 2D kernel density estimation (KDE, the red contour lines) and obtain its value at the location of each of the 115 stars in the sample. Then we tentatively identify the stars in the 1G population as those that have a KDE values that significantly differs from zero - i.e. greater than 5\% of its maximum value - and we highlight them as solid, dark grey points in Figure \ref{pic:extremes_6752}. 

We get that $37\pm 5$ stars belong to the 1G, corresponding to $\sim32.2\pm4.2 \%$, in good agreement with both \cite{milone_2013} and \cite{milone_2017}. The error on these numbers has been estimated via a bootstrapping procedure\footnote{In a nutshell, we made 5000 realizations of the HB in NGC\,6752, identified the 1G stars in each iteration, and took as 1$\sigma$ error the standard deviation of the results.}. For completeness, the histograms in the bottom right panel of Figure \ref{pic:extremes_6752} directly compare the colour distribution of the selected 1G stars with the simulated ones; the red histogram is normalised at the same number of stars in the selected sample. 

Finally, to provide a more informative view of this result, we plot the selected 1G stars together with the other stellar populations along the HB of NGC\,6752 in Figure \ref{pic:tracks_6752}. In the Figure, the 1G stars we previously identified are plotted as the red, open points, together with the zero age horizontal branch (ZAHB) loci of appropriate chemistry (top panel) and the stellar models describing their post ZAHB and post HB evolution (bottom panel). The solid coloured diamonds in the figure highlight the position of the end of the core helium burning.

To study the 2Ge we apply the technique from T20 using the helium mass fraction for the bluest main sequence provided by \cite{milone_2013}. Briefly, we calculate grids of simulated HBs with different parameters, compare them with the photometric data and choose, as best fit, the simulation minimizing the chi-squared distance (hereinafter $\rm \chi^2_d$) between the colour histograms of the data and the simulated stars. In this case our algorithm is set to look for the simulation that best matches the bluest group of stars on the HB.

For the 2Ge of NGC\,6752 the grids have the parameters listed in Table \ref{tab:grids_6752}.
The comparison with the data yields that the best fit simulation is the one with $\rm \mu= 0.285\pm 0.024 M_\odot$ and $\rm \delta= 0.004\pm 0.002 M_\odot$. The errors on these values include different sources of uncertainties. As expected, this value is slightly larger than the one we found previously because here we adopted a slightly lower value of helium enhancement.

The upper panel of Figure \ref{pic:extremes_6752} reports the comparison of the best fit simulation (blue contours) with the data and a typical realisation (solid blue points), while we report the comparison between the colour histograms of the simulation with the selected stars - represented as solid black points in the top panel of the figure - in the lower left panel. The identification is performed in the same way as the 1G case. We identify $24\pm 4$ stars as 2Ge, representing the $\sim20.9\pm 3\%$ of all stars in the branch. As in the case of the 1G stars, the errors have been estimated via bootstrapping.

As in the case of the 1G, we plot the selected 2Ge stars and the models involved as the open blue points and tracks in Figure \ref{pic:tracks_6752}. We find it worth noting that, as shown by the models in the figure, the three stars at $\rm m_{F275W} \sim 14.5$ and $\rm m_{F275W}-m_{F814W}\sim -2.6$ are compatible with being in the post-HB evolutionary phase and should be included in this population, bringing the total number of stars in it to 27 ($\sim 23.5\%$). We report the main parameters of these two populations in Table \ref{tab:par_hb_6752}.

\subsection{Mass loss of the intermediate second generation stars}
\label{sub:intermediate_6752}

\begin{figure}
    \centering
    \includegraphics[width=\columnwidth]{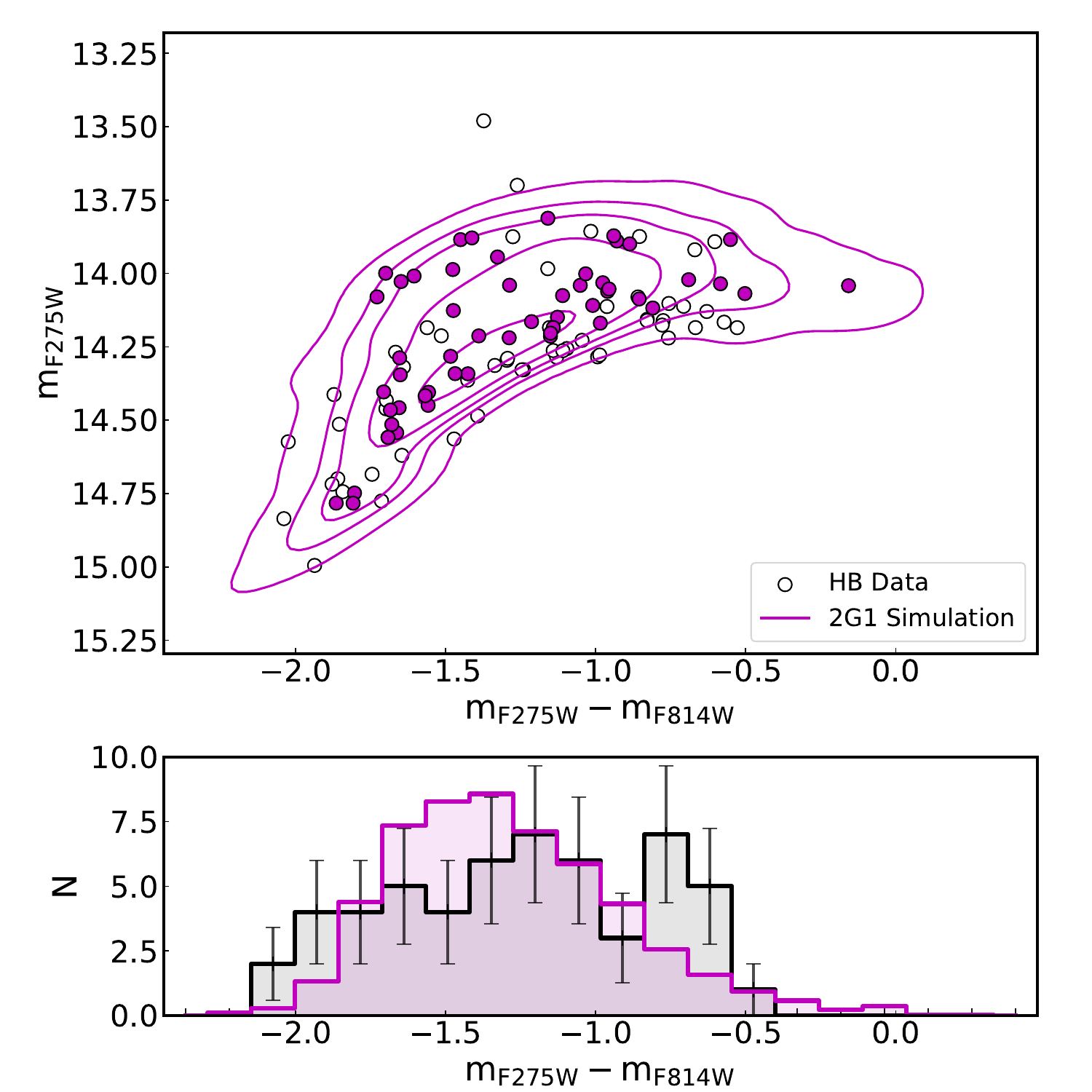}  
    \caption{Contour plot and typical realisation of the best fit simulation for the intermediate stellar population in NGC\,6752. The histograms in the bottom panel compare the colour distributions of both; the magenta one is normalised to the total number of stars in the plot.}
    \label{pic:2gint_6752_wrong}
\end{figure}

\begin{figure}
    \centering
    \includegraphics[width=\columnwidth]{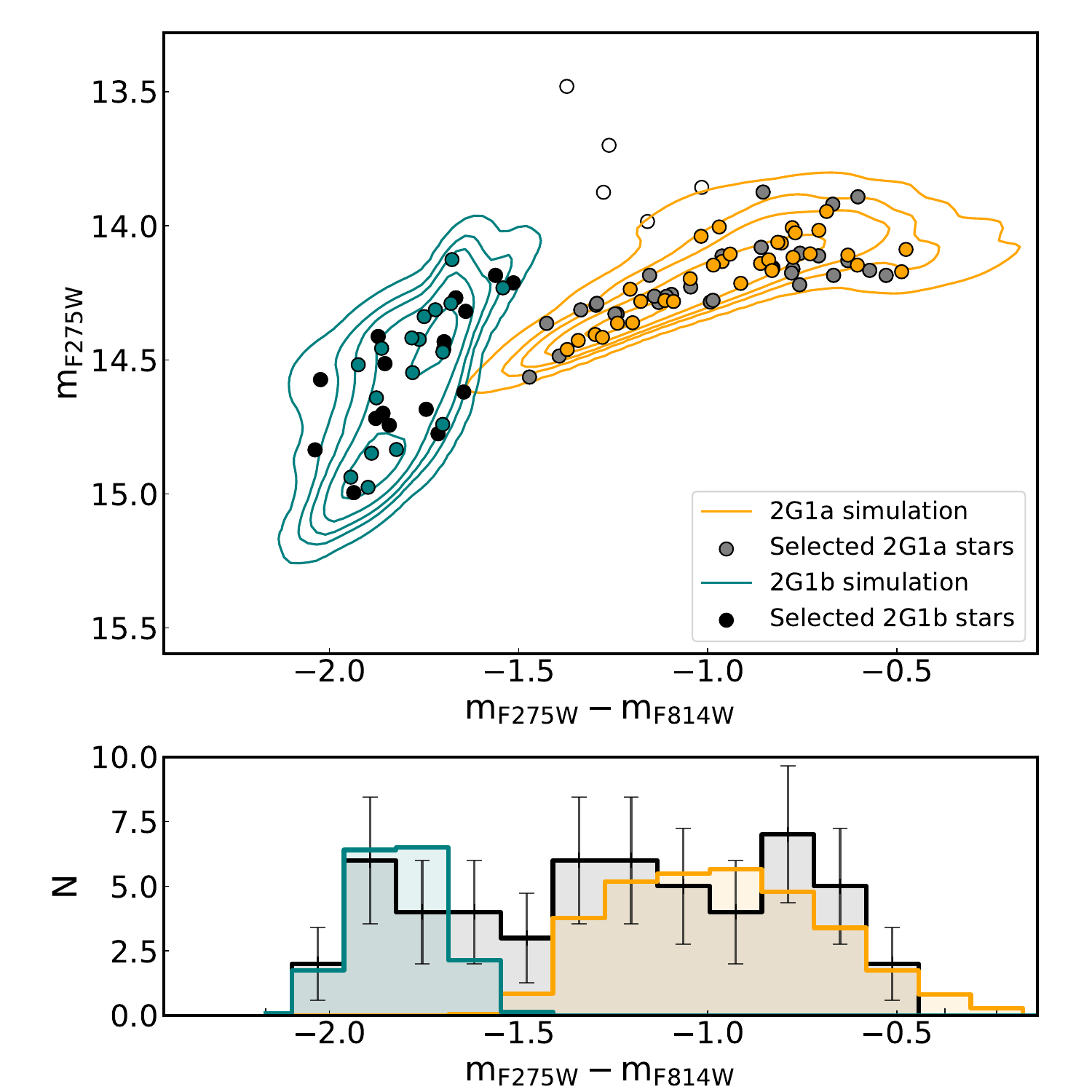}    
    \caption{As Fig.\,\ref{pic:extremes_6752} but for the case where the intermediate HB in NGC\,6752 s described with two individual stellar populations.}
    \label{pic:2gint_6752}
\end{figure}

Here we discuss how we interpret and simulate the stars in the intermediate HB (iHB) and its stellar populations. 

First we remove from the CMD those stars that have been previously identified as 1G and 2Ge. Then, we try to simulate this part of the branch with a single stellar population (see \S\,\ref{sec:data_and_models}). We build the appropriate grids (see Table \ref{tab:grids_6752}) and look for the best fit.
Our procedure identifies the minimum of the $\rm \chi^2_d$ distribution at $\rm \mu= 0.246\, M_\odot$ and  $\rm \delta = 0.017\, M_\odot$. The problem with this result is that, even on a visual inspection, the selected simulation is not a good fit because it does not describe well the peaks we see in the colour distribution of the stars, as demonstrated by the plot in the bottom panel of Figure \ref{pic:2gint_6752_wrong}.
Therefore, we must first make an additional assumption. 

\cite{milone_2013}, using a combination of  UV and optical filters, and \cite{Milone_2019}, using infrared bands, have shown that the intermediate MS is clearly broadened compared to the others, down to the low mass regime, as seen in the IR. This has been interpreted in both occasions as a sign of mixed populations. 
We now associate the reddest group of stars of this part of the HB to the 2G1a, we calculate synthetic HB grids with appropriate parameters (see  Table \ref{tab:grids_6752}) and look for the best fit finding it at $\rm \mu= 0.246\pm0.020\, M_\odot$ and  $\rm \delta = 0.007\pm0.002\, M_\odot$. This simulation is the one represented as the orange contours and solid coloured points in the upper panel of Figure \ref{pic:2gint_6752}. Using the 2D KDE obtained from the best fit simulation we associate 37$\pm$5 stars to this group, giving us a population ratio of $32.2\pm4.3\%$. The selected stars are represented as solid, dark grey points in Figure \ref{pic:2gint_6752}, together with a sample realisation of the simulation.

To study the bluest part of the intermediate branch, which we dub 2G1b, we use the simulation grid described in Table \ref{tab:grids_6752}.
We find that the best fit simulation\footnote{As in previous cases the error estimates on these mass loss values are obtained with the methods by T20, while the error values of the population ratio are obtained via bootstrapping.} is the one with $\rm \mu= 0.270\pm0.021\, M_\odot$ and  $\rm \delta = 0.005\pm0.002\, M_\odot$. This simulation is represented by the teal contour plot and points in the main panel of Figure \ref{pic:2gint_6752}. We associate to the 2G1b 17$\pm$3.6 stars, representing 14.78$\pm$3.1\% of the total stars.

Collectively the two populations host $\sim46\pm5.3\%$ of the total stars on the HB. Furthermore the output average helium abundance of this intermediate population is 0.258. Both values are in good agreement with the results from \cite{milone_2013}. Direct comparison of the histograms of the two best fit simulations and the data is reported in the bottom right panel of Figure \ref{pic:2gint_6752}, which results in a better description than the single population case.Finally it is worth noting that in this CMD there is a group of stars compatible with being post HB of the 2G2 stars, located at $\rm m_{F275W} \sim 14.0 - 13.5$, $\rm m_{F275W}-m_{F814W} \sim -1.2$.  Furthermore the lone star at $\rm m_{F275W} \sim 14.25 - 14.0$, $\rm m_{F275W}-m_{F814W} \sim 0.25$ can be associated to the post HB phase of the 2G1. This is shown by the HB tracks plotted in the bottom panel of Figure \ref{pic:tracks_6752}. The main parameters of these two populations are reported in Table \ref{tab:par_hb_6752}.

\section{The HB of NGC2808}
\label{sec:2808}

In this section we will discuss the detailed analysis of the HB in NGC\,2808.
Initially, we will illustrate its morphology (\S\,\ref{sub:morph_2808}). Afterwards we will discuss the populations found in the rHB (\S\,\ref{sub:2808_rhb}), and then the ones populating its bHB (\S\,\ref{sub:2808_bhb}). 

\subsection{Morphology}
\label{sub:morph_2808}

\begin{figure*}
    \centering
    \includegraphics[width=2.0\columnwidth,trim={0cm 1cm 0cm 0cm}]{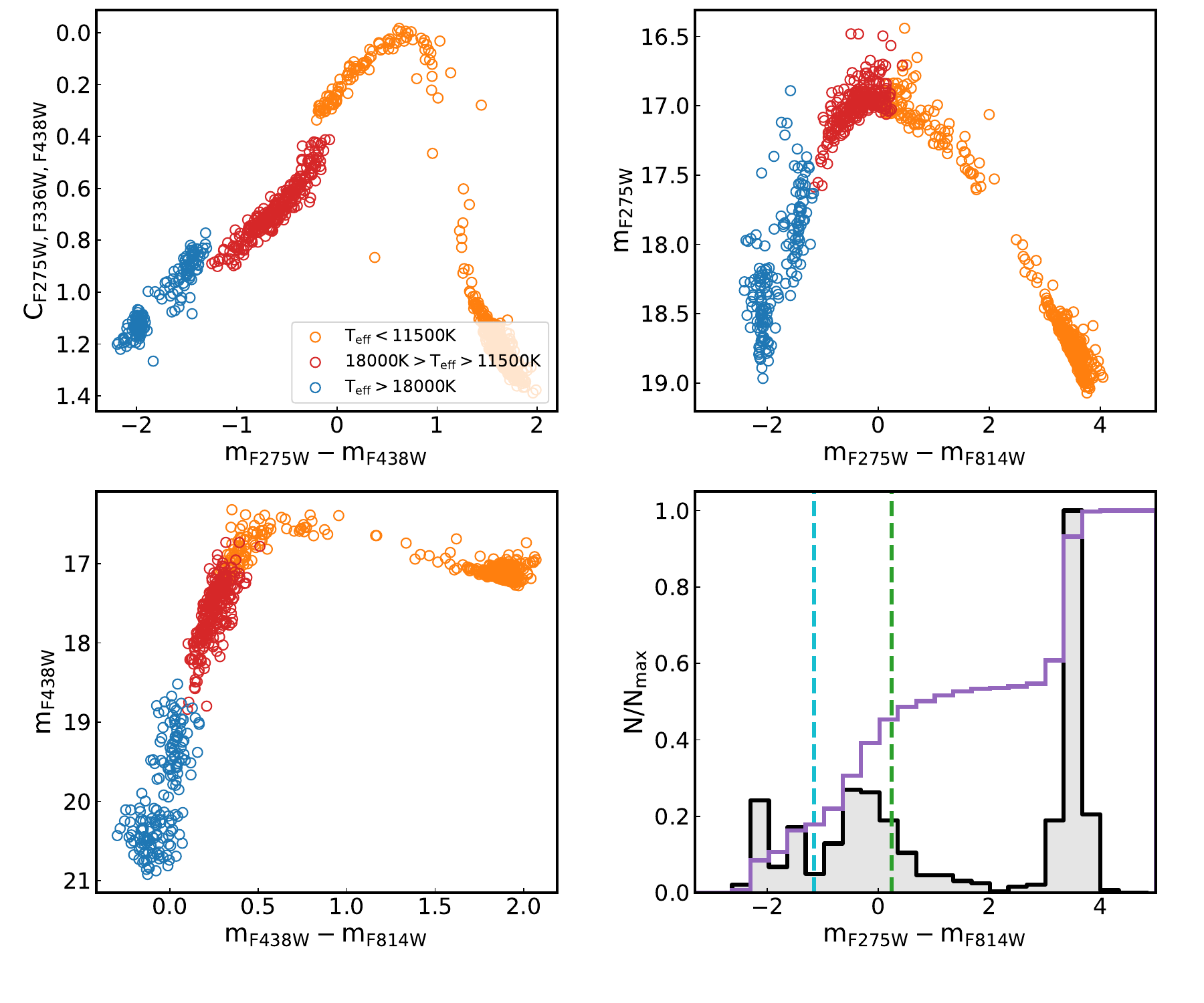}  
    \caption{\textit{Top left panel:} $\rm C_{F275W,F336W,F814W}$ vs $\rm m_{F275W}-m_{F438W}$ pseudo-CMD of the HB stars in NGC\,2808. The stars with T<11500 K, T>18000 K, and 11500 K < T  < 18000 K are marked with orange, blue, and red points to show the location of the G- and M- jumps.\textit{Top right panel:} $\rm m_{F275W}$ vs. $\rm m_{F275W}-m_{F814W}$ CMD of the HB stars in NCG\, 2808. \textit{Bottom right panel:} Histogram of the colour distribution of the photometric data. The cumulative distribution of the points is plotted as the purple profile. The dashed lines in the panel, green and cyan, represent the position in  $\rm m_{F275W}-m_{F438W}$ of the G- and M- jumps, respectively. \textit{Bottom right panel:} $\rm m_{F438W}$  vs. $\rm m_{F438W}-m_{F814W}$ CMD of the HB stars in NCG\, 2808. The colour coding of the points highlight the temperature ranges of the stars where applicable.} 
    \label{pic:morph_2808}
\end{figure*}

Figure \ref{pic:morph_2808} illustrates the morphology of the HB in NGC\,2808 in optical and UV filters. The HB in this cluster hosts $839$ stars populating a large range of colour values and separated into easily identifiable groups, as seen in the CMDs plotted in Figure \ref{pic:morph_2808}.  \cite{milone_2014} found a value of $\sim$0.094 mag for the parameter $\rm L_1$. This classifies NGC\,2808 as M3-like (as in T20).  The IS is easily identifiable in all the diagrams, thus we find that the rHB hosts 362 stars ($\sim 43\%$ of the total) whereas the bHB hosts 465 stars ($\sim53\%$).

The upper left panel of Figure \ref{pic:morph_2808} reports the $\rm C_{F275W,F336W,F438W}$ vs. $\rm m_{F275W}-m_{F438W}$ diagram for this cluster. As expected due to the higher metallicity, the discontinuities on the HB are more evident compared to the NGC\,6752 case. We identify $\sim$580 stars ($\sim 57\%$, orange)  before the G-Jump and $\sim$170 stars ($\sim 17\%$, blue) after the M-jump. Consequentially we find that $\sim 260$ ($\sim 26\%$, red) stars populate the region between the two jumps.

We report the identification of these three ranges of temperature in the other CMDs in Figure \ref{pic:morph_2808}. More specifically by looking at the $\rm m_{F275W}-m_{F814W}$ CMD in the top, right panel in Figure \ref{pic:morph_2808} we see that the position of the G-jump is at $\rm m_{F275W}-m_{F814W}\sim 0.24$ and that M-jump is located at $\rm m_{F275W}-m_{F814W} \sim -1.16$.
To better describe the morphology of this HB, we report, in the bottom right panel of Figure \ref{pic:morph_2808}, the $\rm m_{F275W}-m_{F814W}$ histogram for these HB stars (black) and their cumulative distribution (red). The distribution of stars along the branch has four major peaks. From the left, the first two are in the blue at $\rm m_{F275W}-m_{F814W}\sim -2.12$ and $\sim\,$-1.48 respectively; they host collectively $\sim 18\%$ of the stars in the branch. A third, broader peak is at $\rm m_{F275W}-m_{F814W}\sim -0.29$ and hosts $\sim 35\%$ of stars. Finally the highest peak is at $\rm m_{F275W}-m_{F814W}= 3.5$ and host $\sim 47\%$ of the stars in the HB. In addition the colour location of the two jumps is reported as the green and purple dashed lines, respectively for the G- and M- jumps. We see that, as in the case of NGC\,6752, for the G-jump, no gaps are present at the location of the jumps, reinforcing their independence from the MPs phenomenon.

For completeness we also report the $\rm m_{F438W}$ vs. $\rm m_{F438W}-m_{F814W}$ optical CMD in the lower left panel of Figure \ref{pic:morph_2808}. Similar considerations to those we made for the $\rm m_{F275W}-m_{F814W}$ case can also be made here. We now proceed to study the stellar populations found in this HB,starting from the red side.

\begin{table}
    \centering
    \caption{Parameters of the HB simulation grids used for NGC\,2808. }
    \begin{tabular}{cccc}
        \hline
        \hline
        ID & Y &$\rm \mu_{range}/M_\odot$ &$\rm \delta_{range}/M_\odot$ \\
        \hline
        1G    & 0.250 & 0.03 -- 0.24 & 0.001 -- 0.150 \\  
        2G1   & 0.252 & 0.03 -- 0.24 & 0.001 -- 0.150 \\  
        2G2   & 0.290 & 0.10 -- 0.24 & 0.002 -- 0.030 \\ 
        2Gea  & 0.330 & 0.10 -- 0.25 & 0.001 -- 0.010 \\ 
        2Geb  & 0.360 & 0.10 -- 0.25 & 0.001 -- 0.010 \\ 
        \hline
        \multicolumn{4}{c}{Alternate eHB description}\\
        \hline
        2Gea, 2Geb & 0.340 & 0.10 -- 0.25 & 0.001 -- 0.010 \\  
        \hline
        \hline
    \end{tabular}
    \tablefoot{Columns are: The ID of the population with its value of Y and the ranges of $\rm \mu$ and $\rm \delta$ values. We adopt a step of $\rm 0.003 M_\odot$ and $\rm 0.001 M_\odot$ for the former and the latter.}
    \label{tab:grids_2808}
\end{table}

\begin{table*}
    \centering
    \caption{Parameters of the HB population in NGC\,2808.}
    \begin{tabular}{ccccccccc}
        \hline
        \hline
        ID & Y &N & N/N$_{\rm tot}$& $\rm \mu/M_\odot$&$\rm \delta/M_\odot$&$\rm M_{Tip}/M_\odot$&$\rm \bar{M}_{HB}/M_\odot$&$\rm \Delta\mu/M_\odot$\\
        \hline
        1G   &0.250 & -- & -- &$0.110\pm 0.024$   &$0.009\pm 0.003$   &$0.828$    &$0.718\pm 0.024$   & --\\  
        2G1  &0.252 & -- & -- &$0.126\pm 0.022$   &$0.011\pm 0.003$   &$0.825$    &$0.699\pm 0.023$   &$0.016\pm0.012$\\  
        2G2  &0.290 & $294\pm 15$ & $35.04\pm 1.8\%$ &$0.203\pm 0.023$   &$0.027\pm 0.002$   &$0.773$    &$0.570\pm 0.026$   &$0.093\pm0.014$\\  
        2Gea  &0.330 & $48\pm 7$ &  $5.72\pm 1.0\%$  &$0.226\pm 0.025$   &$0.004\pm 0.001$   &$0.719$    &$0.493\pm 0.025$   &$0.116\pm0.013$\\  
        2Geb  &0.360 & $54\pm 9$ &  $6.51\pm 1.0\%$   &$0.236\pm 0.024$   &$0.008\pm 0.002$   &$0.679$    &$0.443\pm 0.024$   &$0.123\pm0.013$\\  
        \hline
        \multicolumn{7}{c}{Alternate eHB description}\\
        \hline
        2Gea  & 0.340 & $47\pm 7$ &  $5.55\pm 1.0\%$  & $0.213\pm 0.021$   &$0.005\pm 0.001$   &$0.705$    &$0.492\pm 0.021$   &$0.103\pm0.013$\\  
        2Geb  & 0.340 & $53\pm 8$ &  $6.63\pm 1.0\%$   & $0.256\pm 0.024$   &$0.008\pm 0.002$   &$0.705$    &$0.449\pm 0.024$   &$0.146\pm0.013$\\  
        \hline
        \hline
    \end{tabular}
    \tablefoot{Columns are: name of the population (ID), helium abundance (Y), number and number fraction (N and $\rm N/N_{tot}$),  average mass loss ($\mu/M_\odot$), mass loss spread ($\rm \delta/M_\odot$), mass of the star at the Tip of the RGB ($\rm M_{Tip}/M_\odot$), average HB mass for the star in the population ($\rm \bar{M}_{HB}/M_\odot$) and mass loss difference compared to the 1G stars ($\rm \Delta\mu/M_\odot$).}
    \label{tab:par_hb_2808}
\end{table*}

\subsection{Mass loss of the stellar populations in the red HB}
\label{sub:2808_rhb}

\begin{figure*}
    \centering
    \includegraphics[width=2.0\columnwidth]{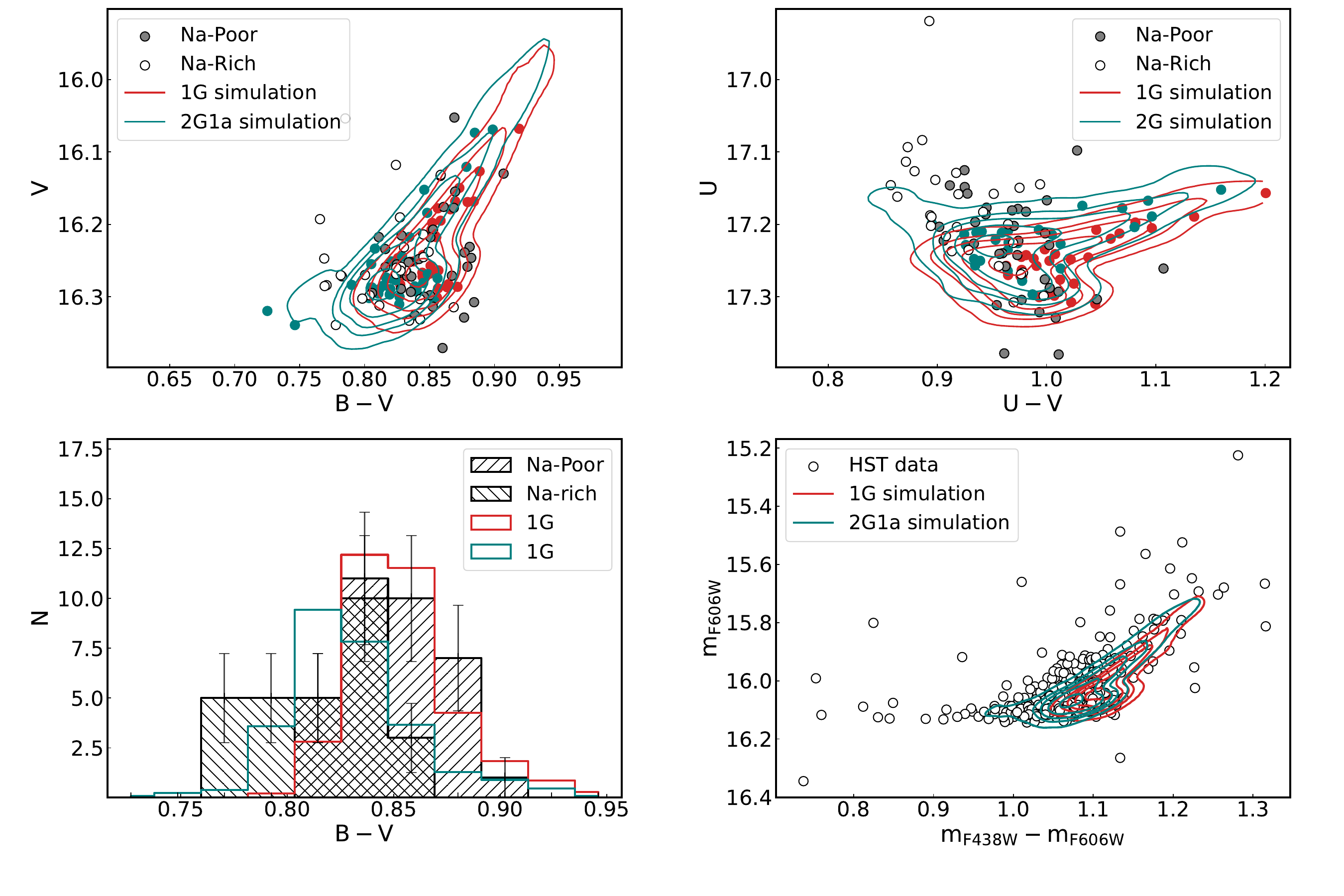}
    \caption{\textit{Top left panel:} Colour magnitude diagram in the V vs. B-V plane of the photometric data of the HB stars in NGC\,2808 from \citet{momany_2004}. The Na-Poor and Na-Rich populations from \citet{marino_2014} are represented with filled and open points, respectively. The red and teal contours  and points represent the best fit simulations for the 1G and the 2G1 in this cluster, respectively, together with a sample realisation. \textit{Bottom left panel:} Histograms of the colour distribution of the stars belonging to the two populations in the data and in the best fit simulations. The histograms of the simulated stars are normalised to the same number of stars we see in the observed sample. \textit{Top right panel:} As the top left panel but for the U vs. U-V CMD. \textit{Bottom right panel:} The contour plot of the best fit simulation for the 1G and 2G1 populations compared with the $\rm m_{F606W}$ vs. $\rm m_{F438W}-m_{F606W}$ CMD for the \textit{HST} photometric catalogue.}
    \label{pic:rhb_2808}
\end{figure*}

\begin{figure}
    \centering
    \includegraphics[width=\columnwidth]{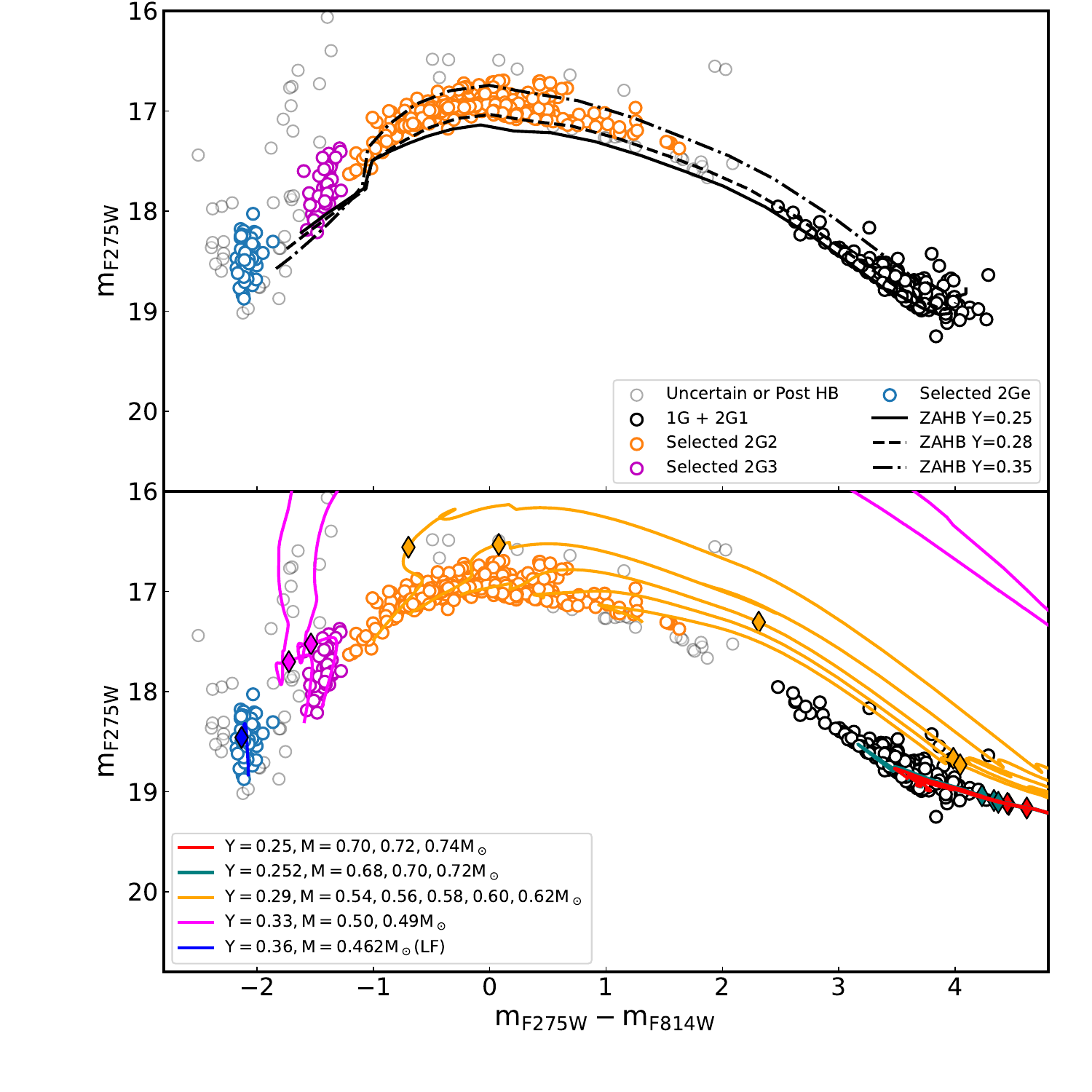}    
    \caption{ \textit{Top panel:} $\rm m_{F275W}$ vs. $\rm m_{F275W}-m_{F814W}$ CMD of the HB in NGC\,2808. Each population we identify is highlighted with the same colour code used in other part of this work. The three lines represents the ZAHB loci of appropriate as in Figure\ref{pic:tracks_6752}.
    \textit{Bottom panel:} As the top panel but we show the stellar tracks involved in the simulation.}
    \label{pic:tracks_2808}
\end{figure}

The rHB of NGC\,2808 is more complex than the one found in NGC\,6752. Indeed, while the latter hosts one population, the results from \cite{marino_2014} show that the former is composed of a mixture of two stellar populations, one more sodium-poor than the other (see their Figure 11). Combining this results with other spectroscopic and spectro-photometric analysis \citep[see e.g.][and references therein]{carretta_2009,milone_2015,marino_2017,Carretta_2018}, it is clear how the low-sodium population is the 1G, accounting for Population A and B, while the high sodium population is part of the least enhanced part of the 2G, accounting for Population C , which here we dub 2G1. This is also corroborated by the information contained in the cumulative distribution of the HB stars plotted in Figure \ref{pic:morph_2808}, as we see there that the rHB contains roughly 47\% of stars, corresponding to the sum of the A, B and C populations.

Therefore, in this more complicated scenario, our method would provide unreliable results because there are mixed populations in the rHB. \cite{milone_2012b} and \cite{Dondoglio2021ApJ...906...76D} showed that in clusters with cool enough rHB stars a special combination of filters is able to separate the stellar population found there. In the case of NGC\,2808 the stars in the rHB are too hot to be effectively separated. We will then use the spectroscopic results from \cite{marino_2014} and the cross-matching they performed with the ground based photometric catalogue from \citet{momany_2004}. Our procedure and results are summarised in Figure \ref{pic:rhb_2808}. 

We produce adequate simulation grids (see Table \ref{tab:grids_2808}) and then compare them with the data to find the best-fit simulations.
We find that the best fit for the 1G is located at $\rm \mu= 0.110 \pm 0.024 M_\odot$ and $\rm \delta= 0.009\pm 0.003M_\odot$, while for the 2G1 the best fit values are $\rm \mu=0.126\pm 0.023M_\odot$ and $\rm \delta= 0.011\pm 0.003 M_\odot$\footnote{The error value includes all the sources already discussed in T20}. We report the best fit simulation as the red and teal contour plots, respectively for the 1G and the 2G1, in the panels of Figure \ref{pic:rhb_2808}. To also show the fit qualitatively, in the lower left panel of Figure \ref{pic:rhb_2808} we report and compare the histograms of the B-V colour distributions for the two populations of stars, as labelled, with their best fit simulations, plotted with the same colour coding adopted for the contour plot in the upper left panel. As in other similar instances throughout this work the series of solid, coloured points represent a sample realisation of the simulations. For completeness the two right panels in Figure \ref{pic:rhb_2808} reproduce the U vs. U-V  (top) and $\rm m_{F606W}$ vs. $\rm m_{F438W}-m_{F606W}$ CMDs (bottom), to show how the best fit simulations look in those planes and bands. As in previous cases, all relevant parameters of the simulated populations are summarized in Table \ref{tab:par_hb_2808}.

Finally we plot the tracks involved in this simulation, together with the ZAHBs of appropriate chemistry, in Figure \ref{pic:tracks_2808}, where we plot the $\rm m_{F275W}$ vs. $\rm m_{F275W}-m_{F814W}$ CMD highlighting the selected stars in each populations. Note that for the 1G and the 2G1, this means highlighting the entirety of the rHB, as we have no means, for now, to separate the population in \textit{HST} bands.
 
\subsection{Mass loss of the stellar populations in the blue HB}
\label{sub:2808_bhb}

In this subsection we discuss the parameters of the stellar populations hosted in the bHB of NGC\,2808. The bHB is the part of the HB located at bluer colour than the IS. In the case of NGC\,2808 it can be divided into two other sections, an iHB, located right after the IS and around the knee of the HB in Figure \ref{pic:morph_2808}, and the eHB, hosting the two bluest groups of stars in the branch. We will discuss the iHB first and then explore the more complex case of the eHB.

\subsubsection{The iHB}
\label{sub:2808_ihb}

\begin{figure}
    \centering
    \includegraphics[width=\columnwidth]{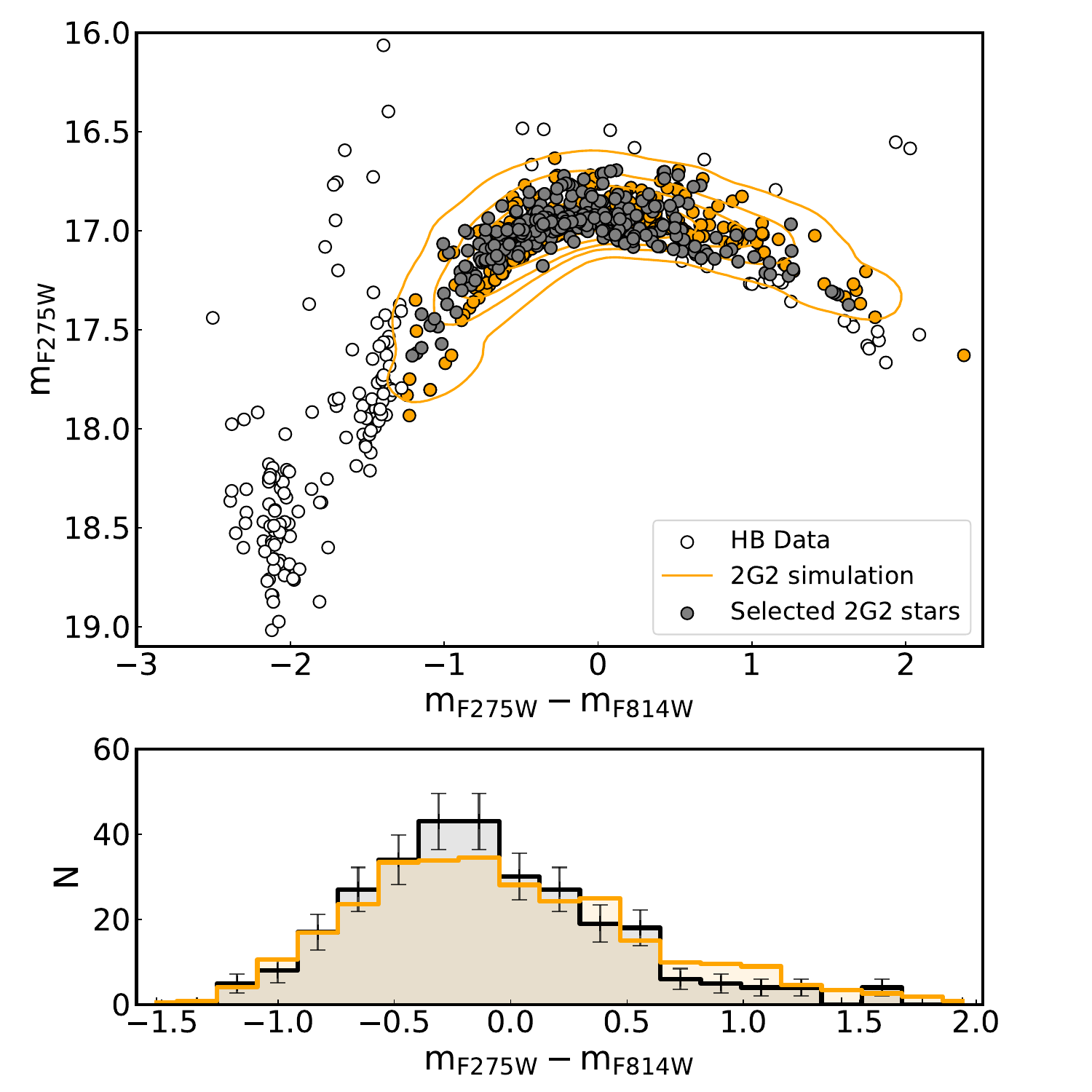}    
    \caption{As Fig. \ref{pic:2gint_6752_wrong} but for the 2G2 stars in NGC\,2808.}
    \label{pic:2808_ihb}
\end{figure}

The iHB in NGC\,2808 is the part of the HB that starts immediately after the blue side of the IS. From the cumulative distribution reported in Figure \ref{pic:morph_2808} we see that this section of the branch hosts $\sim$35\% of the HB stars. The G-jump is located in this section of the HB (see Figure \ref{pic:morph_2808}) but it does not correspond to an actual gap, as the distribution of the stars is continuous both in the $\rm m_{F275W}$ vs. $\rm m_{F275W}-m_{F814W}$ and $\rm m_{F438W}$ vs. $\rm m_{F438W}-m_{F814W}$ CMDs proving again
that the G- and M-jumps are not connected to the presence of different stellar populations. 

From numerical considerations, stemming from the results in \cite{milone_2015} and the spectroscopic indication from \cite{marino_2014}, we can assign this group of stars to Population D. Then, to find the parameters and the extension of this population on the branch, we apply the same method we used in the other part of this work.
We compare an adequate grid of simulations (see  Table \ref{tab:grids_2808}) with the data and find that the simulation minimizing the $\rm \chi^2_d$ is the one with $\rm \mu=  0.203 \pm 0.023 M_\odot$ and $\rm \delta= 0.027\pm 0.003 M_\odot$.

As in other instances in this work, we assign to this population those stars that have a KDE significantly different from zero, finding that the 2G2 hosts $294\pm15$ stars. A direct comparison of the simulation and the data is represented via the CMD and the histograms in Figure \ref{pic:2808_ihb} and \ref{pic:tracks_2808}. In the latter we also show the tracks involved in the simulation and the post HB evolution predicted for these stars. Therefore, the few stars located at  $\rm m_{F275W}\sim 16.5$, and the ones at $m_{F275W}-m_{F814W} \sim 2$,$\rm m_{F275W}\sim 17.0$ are compatible to be stars in the post-HB phase. All other relevant parameters are summarized in Table \ref{tab:par_hb_2808}. 

\subsubsection{The eHB}
\label{sub:2808_ehb}

\begin{figure}
    \centering
    \includegraphics[width=\columnwidth]{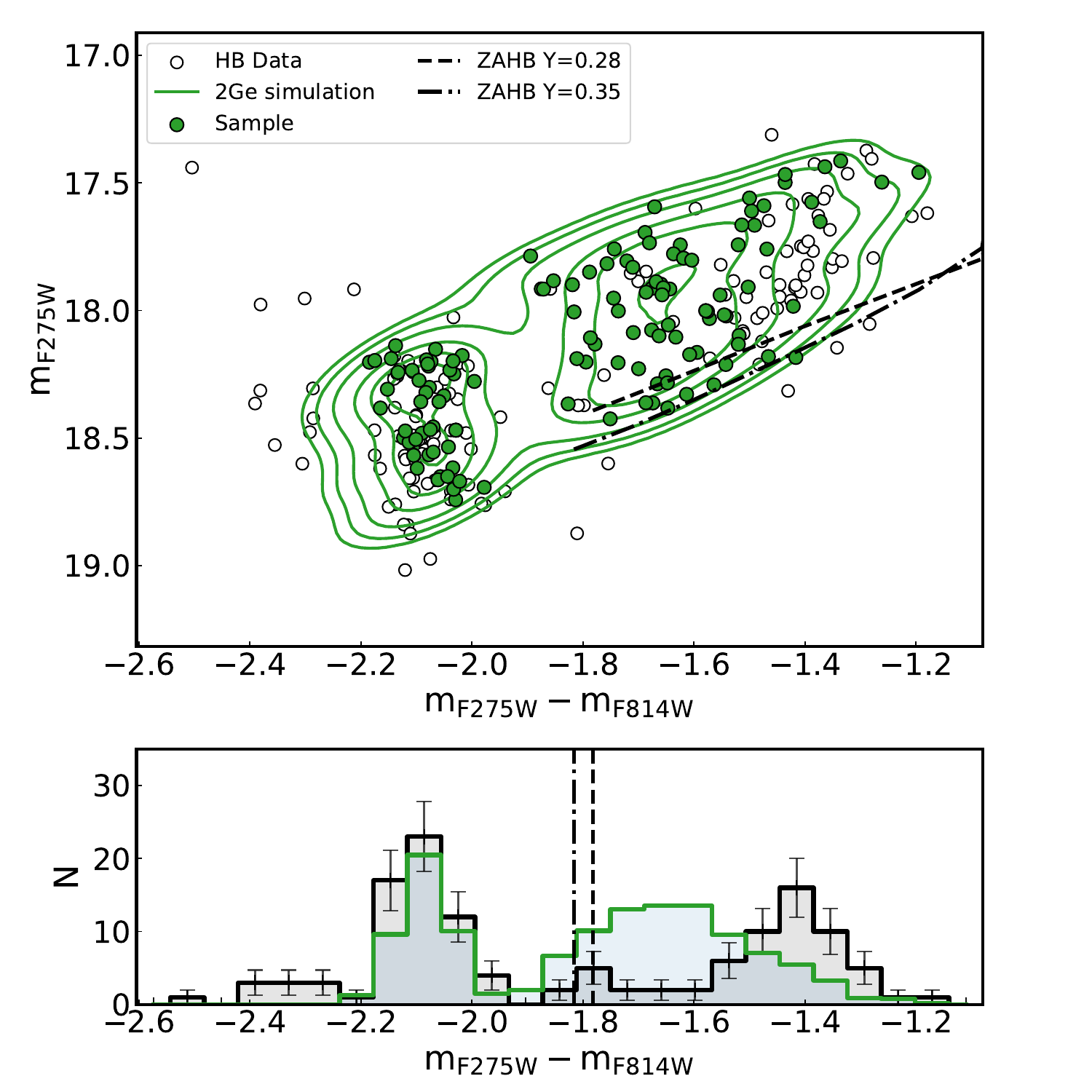}    
    \caption{As Fig.\,\ref{pic:extremes_6752} but for the case of the eHB in NGC\,2808 simulated as an individual group of stars. The dashed and the dot-dashed lines represent the canonical ZAHB of appropriate metallicity with Y=0.28 and Y=0.35, respectively. We also report the end of the canonical HB in the bottom panel.}
    \label{pic:2808_ehb_dantona}
\end{figure}

\begin{figure}
    \centering
    \includegraphics[width=\columnwidth]{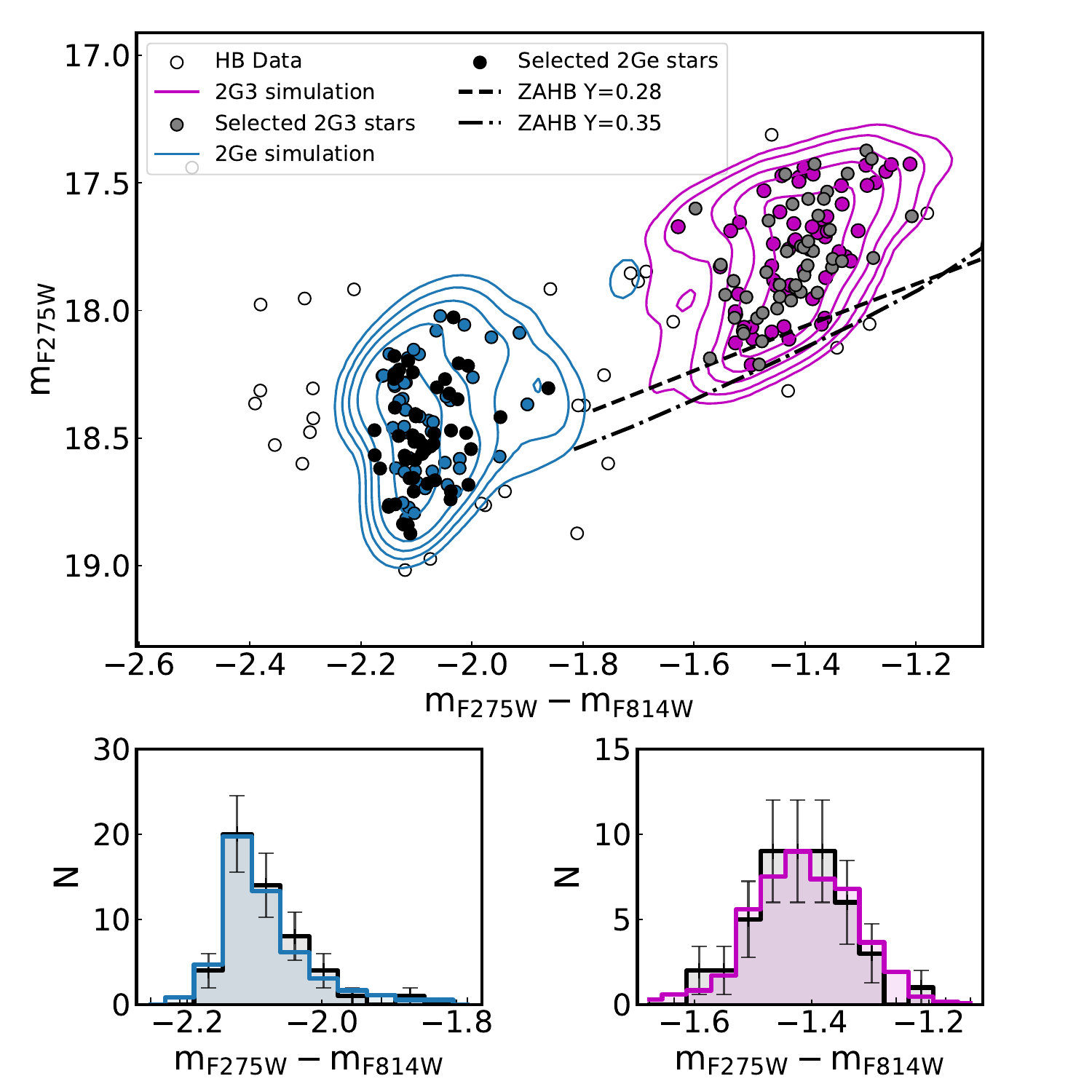}    
    \caption{As Fig.\,\ref{pic:extremes_6752} but for the case of the eHB in NGC\,2808. The dashed and the dot-dashed lines represent the canonical ZAHB of appropriate metallicity with Y=0.28 and Y=0.35, respectively.} 
    \label{pic:2808_ehb}
\end{figure}

The last part of the HB in NGC\,2808, usually referred to eHB, hosts about $\sim$20\% of the total HB stars' number, in two groups of stars well separated in colour. The hottest blue tail, dubbed blue hook \citep[bhk][]{Moehler2004A&A...415..313M} has been generally interpreted as a group of stars suffering an extreme helium flash. The study of the eHB of NGC\,2808 will result in a more complex picture than previously found in the literature, and its complexity could bear consequences on the proposed model of pre--main sequence early disk loss. 

\cite{Lee2005ApJ...621L..57L} and \cite{dantona_2008} first showed that the UV magnitudes and colours of the bhk could only be fit with models having a very high helium abundance in the envelope. \cite{dantona_2008} proposed then that the eHB and the bhk belonged to the same high helium population responsible for the presence of the bluest main sequence in NGC\,2808. The lowest masses in this group would suﬀer a late helium flash with flash-induced mixing (e.g.   \citealt{Castellani1993ApJ...407..649C}, \citealt{Sweigart1997fbs..conf....3S},\citealt{Brown2001ApJ...562..368B},\citealt{Cassisi2003ApJ...582L..43C} or \citealt{tailo_2015} and references therein)  and burn helium in the core at the bhk location. Under this hypothesis, and in the absence of extensive computations following the He--core flash and the flash mixing event, they could parametrize the limiting mass between the standard flash and the flash-mixing in order to describe the eHB and the bhk as a single group. This interpretation was indeed guided by the recent,  at the time, determination of the triple main sequence \citep{dantona_2005, piotto_2007} in this cluster.

We now can take advantage of our new stellar models, which include the flash--mixing explicitly computed in \citet{tailo_2015} and T20. For Y=0.35 the standard evolution extends down to a mass of M=0.475\Msun, while the smallest mass which can populate the EHB is M$\sim$0.49\Msun: there is then a mass range 0.475$<$M/\Msun$<$0.490 which is not present in the data, but would result in the eHB distribution if we adopt a single helium content and a continuous mass loss distribution for the two groups of stars (see Figure \ref{pic:2808_ehb_dantona}). Thus our interpretation of the hot end of the HB in NGC\,2808 needs to become more complicated than in previous approaches\footnote{We warn that a different modelling of the helium flash might in future provide results that account for this 0.02\Msun\ difference in the beginning of the flash mixing event, and change the interpretation again.}.

A first possibility is that Population E is actually composed of two sub populations , we dub 2Gea and 2Geb, the average helium values of which is the one estimated by \citet[][0.34]{milone_2015} that then populate the two clumps observed in the eHB (the EBT1 and EBT2, \citealt{bedin2000}). This interpretation is corroborated by the results by \cite{milone_2012}. In their study, Milone and collaborators found that the bluest main sequence in NGC\,2808 has a higher colour dispersion. This larger spread may indicate, as in the case of NGC\,6752, the presence of more than one stellar population with helium values close enough to merge but, at the same time, different enough to populate different part of the HB. Hence we need to make an additional assumption and assign helium abundance values to these two sub populations. We assign Y=0.33 and Y=0.36, respectively of the 2Gea and the 2Geb. The two groups hosts more or less the same number of stars and these two values give back the Y=0.34 found by \citet{milone_2015} once averaged. In this description the latter of the two groups undergoes the process of late-flash mixing and popoulates the bhk
 We produce two simulation grids that differ only in helium content (see Table \ref{tab:grids_2808}) and compare them with the data, finding that the red side of the eHB (hosting the 2Gea) is best fit by the simulation with Y=0.33, $\rm \mu= 0.226 \pm 0.025 M_\odot$ and $\rm \delta= 0.004\pm 0.001 M_\odot$; on the other hand the bluest group (we identify as the 2Geb) is best described by the simulation with Y=0.36, $\rm \mu=0.236 \pm 0.024 M_\odot$ and $\rm \delta= 0.008\pm 0.002 M_\odot$. 
 The results are described in Figure \ref{pic:2808_ehb} where we plot the two simulations with magenta and blue contour lines, respectively for the 2Gea and the 2Geb. With the same method used elsewhere in this work, we find $47\pm6.9$ and $53\pm7.9$ stars. For completeness, the comparison of the histograms is reported in the bottom panels and follows the same colour coding of the contour plots. The other relevant parameters of these populations are summarized in Table \ref{tab:par_hb_2808}.

An alternative description of the eHB in this cluster can be obtained if we consider the two groups as part of a single population. In this description both groups have the same Y and suffered different amounts of total mass loss, and the group that suffered the most populated the bhk (as suggested by the \citealt{tailo_2015} scenario for the bhk in $\omega$ Centauri). To achieve this description we compare the two parts of the eHB with a grid of model HBs having Y=0.34 and the same range of $\rm \mu$ and $\rm \delta$ of the previous case. 
The simulations that best fit the two groups are the ones corresponding to $\rm \mu= 0.213 \pm 0.021 M_\odot$ with $\rm \delta= 0.005\pm 0.001 M_\odot$, and $\rm \mu= 0.256 \pm 0.024 M_\odot$ with $\rm \delta= 0.008\pm 0.002 M_\odot$,  respectively for the 2Gea and 2Geb. The other parameters of these two stellar populations are summarized in Table \ref{tab:grids_2808} and \ref{tab:par_hb_2808}.  It is worth noting that our late flash model for the most enhanced chemistry has a mass of $\sim 0.46$\Msun, while the value of $\mu$ of the best fit simulation gives a lower average mass: this implies that the peak of the mass loss for this population is in an interval that mainly produces helium white dwarfs, maybe explaining some of the missing AGB stars noted in \cite{marino_2017}. This is in agreement with our previous fidngs about $\omega$ Cen and NGC\,2419 \citep{dicriscienzo_2015,tailo_2015}. From the track plotted in the lower panel of Figure \ref{pic:tracks_2808}, we se that the stars located at at $\rm m_{F275W}-m_{F814W}< -1.5$ and $\rm m_{F275W}-m_{F814W}< -2.2$  are compatible with being the post helium burning stars corresponding the two parts of the eHB.

\section{Discussion}
\label{sec:disc}

\begin{figure}
    \centering
    \includegraphics[width=\columnwidth]{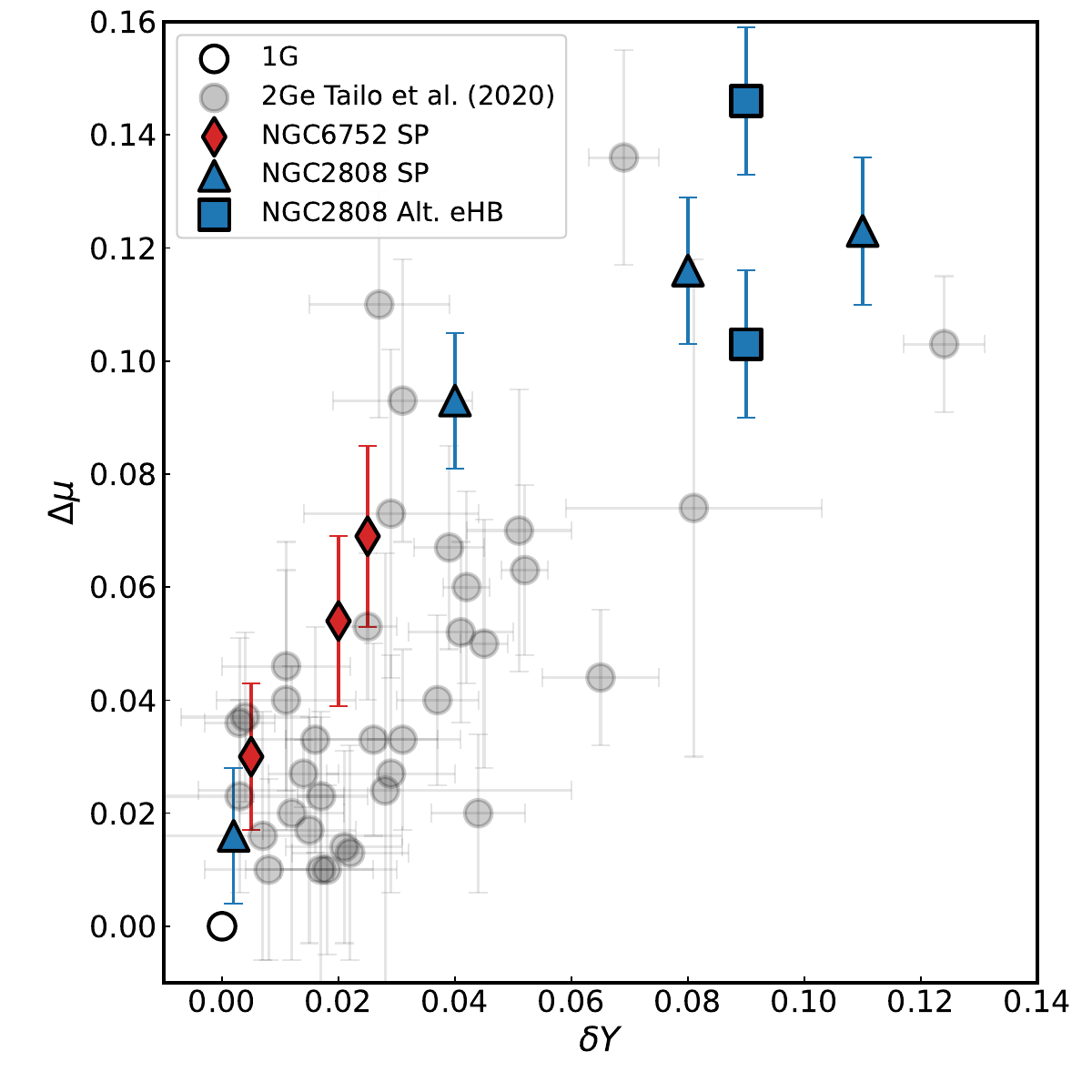}    
    \caption{Mass loss increase ($\rm \Delta \mu$) vs. helium enhancement ($\rm \Delta Y$) of the stellar populations studied in this work. The two series of points refer to the stellar populations (SP) in NGC\,6752 (red diamonds) and NGC\,2808 (blue triangles). The two blue squares in the figure refer to the alternate description of the eHB in NGC\,2808. In the figure the data from \citet{tailo_2020} are also reported.}
    \label{pic:mloss_diff}
\end{figure}

In previous sections and subsections we analysed the stellar populations found along the HB of NGC\,6752 and NGC\,2808, while at the same time breaking the traditional parameter degeneracy associated with these stars using helium abundance estimates from the literature (see \S\,\ref{sec:data_and_models}). Our findings suggest that the total, integrated mass loss suffered by these stars during the ascent of the RGB needs to be significantly increased in the stellar populations past the 1G.

More specifically, in NGC\,6752  the minimum mass loss associated to the 1G is $0.216\,M_\odot$ while the maximum value for the mass loss is $0.285\,M_\odot$ for the 2Ge. Our values of $\mu$ and $\rm \bar{M}_{HB}$ are in agreement, within their errors, with the ones in \cite{Cassisi2014A&A...571A..81C}. In NGC\,2808 the minimum mass loss, associated to the 1G, is $0.113\,M_\odot$ while the maximum value of the mass loss is $0.226\,M_\odot$ or $0.256\,M_\odot$ depending on the description adopted to describe the eHB in that cluster. Therefore a non negligible increase in mass loss is necessary to describe each stellar population on the HB. The exact value of the needed increase is reported in Tables \ref{tab:par_hb_6752} and \ref{tab:par_hb_2808}. The errors reported in the table have been obtained with the method from \citet{tailo_2020}.

Figure \ref{pic:mloss_diff} reports the increase of mass loss ($\rm \Delta \mu$) vs. the helium enhancement ($\rm \delta Y$) of the stellar populations analysed in this paper. The red diamonds are for the stellar populations in NGC\,6752 and the blue triangles are for the NGC\,2808 ones. The two blue squares are for the alternative description of NGC\,2808 eHB we provided in \S\,\ref{sub:2808_ehb}. In the Figure we also plot the data for the 2Ge from T20 as grey, semi-transparent points. We clearly see that all the analysed stellar populations describe a clear $\rm \delta \mu$ vs. $\rm \delta Y$ relation, compatible with the one described by the points from T20. In other words, we find that the populations having ``intermediate" helium abundances in the two clusters here examined behave like the populations with similar helium abundances in many other clusters. 

This finding reinforces the tentative explanation advanced in T20 (see their Fig.\,18) to explain the results, namely that the increase in mass loss is the consequence of the physical environment in which populations with different helium formed. In the context of the standard ``AGB model", the 2Ge forms at higher density than the 1G, in the very compact core of the GC, where the pure winds of the massive AGBs, with the maximum values of $\rm \delta Y$, collect \citep{dercole_2008}. The intermediate populations form, instead, by mixing of the AGB ejecta with re--accreting pristine gas, resulting in intermediate values of Y, and form in a lower density environment than the 2Ge, as confirmed in the hydrodynamical simulations by \cite{calura_2019}. 
The denser the environment, the earlier the destruction of the protostellar accretion disk will be: this will lock the stellar rotation with the disk rotation at higher velocities and make the acceleration of the stellar rotation more effective due to the conservation of angular momentum in the contracting pre-main sequence star.
If the early higher rotation rate implies a larger average mass loss rate, 
the observed relation of Figure \ref{pic:mloss_diff} results to be naturally explained as the fossil trace of the formation mechanism.

A complication in our result is the ambiguous way in which we have reproduced the eHB plus bhk stars in NGC\,2808\footnote{This problem did not emerge in T20, where only the most extreme group, the bhk stars, had been examined.} If we trust the models, these two groups have either a difference in helium content or a difference in mass loss. The different helium content by $\Delta$Y$\sim$0.02 could imply an inhomogeneity in the helium content of the stars forming from the pure ejecta of the most massive AGBs or super--AGBs and requires careful investigation of the AGB model results. An abrupt difference in mass loss rate could be explained if the birth location of the stars within the cooling flow has a further effect on the early disk loss.

\section{Summary}
\label{sec:concl}

In the present work, we combined high-precision photometry from the UV Legacy survey \citep[][]{piotto_2015,nardiello_2018} of NGC\,6752 and NGC\,2808 with stellar population models to characterize the intermediate stellar populations hosted in their HB. To constrain the helium abundance in each stellar population we used the helium abundances inferred by \citet[][for NGC\,6752]{milone_2013} and \citet[][for NGC\,2808]{milone_2015}. 
The analysis of the HB population has been carried out following the method developed by T20 and T21. We summarized the parameters we derived for each stellar populations in Tables \ref{tab:par_hb_6752} and \ref{tab:par_hb_6752} . Our main results can be summarized as follows:

\begin{enumerate}[(i)]
    \item In NGC\,6752 we divided the HB into four different stellar populations. The values of integrated mass loss we found in our study range from $0.216\,M_\odot$ for the 1G \citep[from][]{tailo_2020} to $0.280\,M_\odot$ for the 2Ge. The other two intermediate populations have values of mass loss located in between these two: $0.246\,M_\odot$ for the 2G1 and $0.270\,M_\odot$ for the 2G2. We report all the details of these stellar populations in Table \ref{tab:par_hb_6752}. 
    \item In NGC\,2808, the description is more complex. We identified five stellar populations and the integrated mass loss our algorithm associated with each of them varies from $\rm 0.113\, M_\odot$, for the 1G to $\rm 0.236\, M_\odot$ (or $\rm 0.256\, M_\odot$ depending on the description adopted) for the 2Ge, populating the bhk of the HB. For the other intermediate populations, corresponding to the different sub populations of the 2G, we found integrated mass loss values of $\rm 0.126\, M_\odot$, $0.203\, M_\odot$ and $0.226\, M_\odot$, as detailed in Table \ref{tab:par_hb_2808}.
    \item Our combined results show that the amount of mass loss correlates with the helium content of the stellar populations. In particular, the pattern of increased mass loss is consistent with the behavior described in T20. We interpret this as the signature of the formation mechanism of the MPs.    
\end{enumerate}

\begin{acknowledgements}
This work has received funding from
"PRIN 2022 2022MMEB9W - {\it Understanding the formation of globular clusters with their multiple stellar generations}" (PI Anna F.\,Marino),
from INAF Research GTO-Grant Normal RSN2-1.05.12.05.10 -  (ref. Anna F. Marino) of the "Bando INAF per il Finanziamento della Ricerca Fondamentale 2022",  and from the European Union’s Horizon 2020 research and innovation programme under the Marie Skłodowska-Curie Grant Agreement No. 101034319 and from the European Union – NextGenerationEU (beneficiary: T. Ziliotto).
SJ acknowledges support from the NRF of Korea (2022R1A2C3002992, 2022R1A6A1A03053472).

\end{acknowledgements}

%
  \bibliographystyle{aa} 
  \bibliography{HB_intermediate} 

\begin{thebibliography}{82}
\expandafter\ifx\csname natexlab\endcsname\relax\def\natexlab#1{#1}\fi

\bibitem[{{Armitage} \& {Clarke}(1996)}]{Armitage1996MNRAS.280..458A}
{Armitage}, P.~J. \& {Clarke}, C.~J. 1996, \mnras, 280, 458

\bibitem[{{Bastian} \& {Lardo}(2018)}]{bastian_2018}
{Bastian}, N. \& {Lardo}, C. 2018, \araa, 56, 83

\bibitem[{{Baumgardt} \& {Hilker}(2018)}]{baumgardt_2018}
{Baumgardt}, H. \& {Hilker}, M. 2018, \mnras, 478, 1520

\bibitem[{{Baumgardt} {et~al.}(2019){Baumgardt}, {Hilker}, {Sollima}, \&
  {Bellini}}]{baumgardt_2019}
{Baumgardt}, H., {Hilker}, M., {Sollima}, A., \& {Bellini}, A. 2019, \mnras,
  482, 5138

\bibitem[{{Bedin} {et~al.}(2000){Bedin}, {Piotto}, {Zoccali}, {Stetson},
  {Saviane}, {Cassisi}, \& {Bono}}]{bedin2000}
{Bedin}, L.~R., {Piotto}, G., {Zoccali}, M., {et~al.} 2000, \aap, 363, 159

\bibitem[{{Bouvier} {et~al.}(1997){Bouvier}, {Forestini}, \&
  {Allain}}]{Bouvier1997A&A...326.1023B}
{Bouvier}, J., {Forestini}, M., \& {Allain}, S. 1997, \aap, 326, 1023

\bibitem[{{Brown} {et~al.}(2016){Brown}, {Cassisi}, {D'Antona}, {Salaris},
  {Milone}, {Dalessandro}, {Piotto}, {Renzini}, {Sweigart}, {Bellini},
  {Ortolani}, {Sarajedini}, {Aparicio}, {Bedin}, {Anderson}, {Pietrinferni}, \&
  {Nardiello}}]{brown_2016}
{Brown}, T.~M., {Cassisi}, S., {D'Antona}, F., {et~al.} 2016, \apj, 822, 44

\bibitem[{{Brown} {et~al.}(2001){Brown}, {Sweigart}, {Lanz}, {Landsman}, \&
  {Hubeny}}]{Brown2001ApJ...562..368B}
{Brown}, T.~M., {Sweigart}, A.~V., {Lanz}, T., {Landsman}, W.~B., \& {Hubeny},
  I. 2001, \apj, 562, 368

\bibitem[{{Brown} {et~al.}(2017){Brown}, {Taylor}, {Cassisi}, {Sweigart},
  {Bellini}, {Bedin}, {Salaris}, {Renzini}, \& {Dalessandro}}]{brown_2017}
{Brown}, T.~M., {Taylor}, J.~M., {Cassisi}, S., {et~al.} 2017, \apj, 851, 118

\bibitem[{{Calura} {et~al.}(2019){Calura}, {D'Ercole}, {Vesperini}, {Vanzella},
  \& {Sollima}}]{calura_2019}
{Calura}, F., {D'Ercole}, A., {Vesperini}, E., {Vanzella}, E., \& {Sollima}, A.
  2019, \mnras, 489, 3269

\bibitem[{Carretta(2015)}]{carretta_2015}
Carretta, E. 2015, \apj, 810, 148

\bibitem[{{Carretta} {et~al.}(2009){Carretta}, {Bragaglia}, {Gratton}, \&
  {Lucatello}}]{carretta_2009}
{Carretta}, E., {Bragaglia}, A., {Gratton}, R., \& {Lucatello}, S. 2009, \aap,
  505, 139

\bibitem[{{Carretta} {et~al.}(2010){Carretta}, {Bragaglia}, {Gratton},
  {Recio-Blanco}, {Lucatello}, {D'Orazi}, \& {Cassisi}}]{Carretta2010}
{Carretta}, E., {Bragaglia}, A., {Gratton}, R.~G., {et~al.} 2010, \aap, 516,
  A55

\bibitem[{{Carretta} {et~al.}(2018){Carretta}, {Bragaglia}, {Lucatello},
  {Gratton}, {D'Orazi}, \& {Sollima}}]{Carretta_2018}
{Carretta}, E., {Bragaglia}, A., {Lucatello}, S., {et~al.} 2018, \aap, 615, A17

\bibitem[{{Carretta} {et~al.}(2011){Carretta}, {Lucatello}, {Gratton},
  {Bragaglia}, \& {D'Orazi}}]{carretta2011}
{Carretta}, E., {Lucatello}, S., {Gratton}, R.~G., {Bragaglia}, A., \&
  {D'Orazi}, V. 2011, \aap, 533, A69

\bibitem[{{Cassisi} {et~al.}(2014){Cassisi}, {Salaris}, {Pietrinferni}, {Vink},
  \& {Monelli}}]{Cassisi2014A&A...571A..81C}
{Cassisi}, S., {Salaris}, M., {Pietrinferni}, A., {Vink}, J.~S., \& {Monelli},
  M. 2014, \aap, 571, A81

\bibitem[{{Cassisi} {et~al.}(2003){Cassisi}, {Schlattl}, {Salaris}, \&
  {Weiss}}]{Cassisi2003ApJ...582L..43C}
{Cassisi}, S., {Schlattl}, H., {Salaris}, M., \& {Weiss}, A. 2003, \apjl, 582,
  L43

\bibitem[{{Castellani} \& {Castellani}(1993)}]{Castellani1993ApJ...407..649C}
{Castellani}, M. \& {Castellani}, V. 1993, \apj, 407, 649

\bibitem[{{Catelan} {et~al.}(1998){Catelan}, {Borissova}, {Sweigart}, \&
  {Spassova}}]{catelan1998}
{Catelan}, M., {Borissova}, J., {Sweigart}, A.~V., \& {Spassova}, N. 1998,
  \apj, 494, 265

\bibitem[{{Dalessandro} {et~al.}(2011){Dalessandro}, {Salaris}, {Ferraro},
  {Cassisi}, {Lanzoni}, {Rood}, {Fusi Pecci}, \& {Sabbi}}]{dalessandro_2011}
{Dalessandro}, E., {Salaris}, M., {Ferraro}, F.~R., {et~al.} 2011, \mnras, 410,
  694

\bibitem[{{Dalessandro} {et~al.}(2013){Dalessandro}, {Salaris}, {Ferraro},
  {Mucciarelli}, \& {Cassisi}}]{dalessandro_2013}
{Dalessandro}, E., {Salaris}, M., {Ferraro}, F.~R., {Mucciarelli}, A., \&
  {Cassisi}, S. 2013, \mnras, 430, 459

\bibitem[{{D'Antona} {et~al.}(2005){D'Antona}, {Bellazzini}, {Caloi}, {Pecci},
  {Galleti}, \& {Rood}}]{dantona_2005}
{D'Antona}, F., {Bellazzini}, M., {Caloi}, V., {et~al.} 2005, \apj, 631, 868

\bibitem[{{D'Antona} \& {Caloi}(2008)}]{dantona_2008}
{D'Antona}, F. \& {Caloi}, V. 2008, \mnras, 390, 693

\bibitem[{{D'Antona} {et~al.}(2002){D'Antona}, {Caloi}, {Montalb{\'a}n},
  {Ventura}, \& {Gratton}}]{dantona_2002}
{D'Antona}, F., {Caloi}, V., {Montalb{\'a}n}, J., {Ventura}, P., \& {Gratton},
  R. 2002, \aap, 395, 69

\bibitem[{{D'Ercole} {et~al.}(2008){D'Ercole}, {Vesperini}, {D'Antona},
  {McMillan}, \& {Recchi}}]{dercole_2008}
{D'Ercole}, A., {Vesperini}, E., {D'Antona}, F., {McMillan}, S.~L.~W., \&
  {Recchi}, S. 2008, \mnras, 391, 825

\bibitem[{{Di Criscienzo} {et~al.}(2015){Di Criscienzo}, {Tailo}, {Milone},
  {D'Antona}, {Ventura}, {Dotter}, \& {Brocato}}]{dicriscienzo_2015}
{Di Criscienzo}, M., {Tailo}, M., {Milone}, A.~P., {et~al.} 2015, \mnras, 446,
  1469

\bibitem[{{Dondoglio} {et~al.}(2021){Dondoglio}, {Milone}, {Lagioia}, {Marino},
  {Tailo}, {Cordoni}, {Jang}, \& {Carlos}}]{Dondoglio2021ApJ...906...76D}
{Dondoglio}, E., {Milone}, A.~P., {Lagioia}, E.~P., {et~al.} 2021, \apj, 906,
  76

\bibitem[{{Dotter} {et~al.}(2011){Dotter}, {Sarajedini}, \&
  {Anderson}}]{dotter_2011}
{Dotter}, A., {Sarajedini}, A., \& {Anderson}, J. 2011, \apj, 738, 74

\bibitem[{{Dotter} {et~al.}(2010){Dotter}, {Sarajedini}, {Anderson},
  {Aparicio}, {Bedin}, {Chaboyer}, {Majewski}, {Mar{\'\i}n-Franch}, {Milone},
  {Paust}, {Piotto}, {Reid}, {Rosenberg}, \& {Siegel}}]{dotter_2010}
{Dotter}, A., {Sarajedini}, A., {Anderson}, J., {et~al.} 2010, \apj, 708, 698

\bibitem[{{Faulkner}(1966)}]{faulkner1966}
{Faulkner}, J. 1966, \apj, 144, 978

\bibitem[{{Fusi Pecci} \& {Bellazzini}(1997)}]{fpb1997}
{Fusi Pecci}, F. \& {Bellazzini}, M. 1997, in The Third Conference on Faint
  Blue Stars, ed. A.~G.~D. {Philip}, J.~{Liebert}, R.~{Saffer}, \& D.~S.
  {Hayes}, 255

\bibitem[{{Fusi Pecci} {et~al.}(1993){Fusi Pecci}, {Ferraro}, {Bellazzini},
  {Djorgovski}, {Piotto}, \& {Buonanno}}]{fusipecci1993}
{Fusi Pecci}, F., {Ferraro}, F.~R., {Bellazzini}, M., {et~al.} 1993, \aj, 105,
  1145

\bibitem[{{Gratton} {et~al.}(2019){Gratton}, {Bragaglia}, {Carretta},
  {D'Orazi}, {Lucatello}, \& {Sollima}}]{Gratton_2019}
{Gratton}, R., {Bragaglia}, A., {Carretta}, E., {et~al.} 2019, \aapr, 27, 8

\bibitem[{{Grundahl} {et~al.}(1999){Grundahl}, {Catelan}, {Landsman},
  {Stetson}, \& {Andersen}}]{grundahl_1999}
{Grundahl}, F., {Catelan}, M., {Landsman}, W.~B., {Stetson}, P.~B., \&
  {Andersen}, M.~I. 1999, \apj, 524, 242

\bibitem[{{H{\"a}rm} \& {Schwarzschild}(1964)}]{hs1964}
{H{\"a}rm}, H. \& {Schwarzschild}, M. 1964, \apj, 139, 594

\bibitem[{{Harris}(1996)}]{harris_1996}
{Harris}, W.~E. 1996, \aj, 112, 1487

\bibitem[{{Harris}(2010)}]{harris_2010}
{Harris}, W.~E. 2010, ArXiv e-prints [\eprint[arXiv]{1012.3224}]

\bibitem[{{Iben} \& {Rood}(1970)}]{ir1970}
{Iben}, Jr., I. \& {Rood}, R.~T. 1970, \apj, 161, 587

\bibitem[{{Lee} {et~al.}(2005){Lee}, {Joo}, {Han}, {Chung}, {Ree}, {Sohn},
  {Kim}, {Yoon}, {Yi}, \& {Demarque}}]{Lee2005ApJ...621L..57L}
{Lee}, Y.-W., {Joo}, S.-J., {Han}, S.-I., {et~al.} 2005, \apjl, 621, L57

\bibitem[{{Legnardi} {et~al.}(2022){Legnardi}, {Milone}, {Armillotta},
  {Marino}, {Cordoni}, {Renzini}, {Vesperini}, {D'Antona}, {McKenzie}, {Yong},
  {Dondoglio}, {Lagioia}, {Carlos}, {Tailo}, {Jang}, \&
  {Mohandasan}}]{Legnardi_2022}
{Legnardi}, M.~V., {Milone}, A.~P., {Armillotta}, L., {et~al.} 2022, \mnras,
  513, 735

\bibitem[{{Mar{\'\i}n-Franch} {et~al.}(2009){Mar{\'\i}n-Franch}, {Aparicio},
  {Piotto}, {Rosenberg}, {Chaboyer}, {Sarajedini}, {Siegel}, {Anderson},
  {Bedin}, {Dotter}, {Hempel}, {King}, {Majewski}, {Milone}, {Paust}, \&
  {Reid}}]{marinf_2009}
{Mar{\'\i}n-Franch}, A., {Aparicio}, A., {Piotto}, G., {et~al.} 2009, \apj,
  694, 1498

\bibitem[{{Marino} {et~al.}(2009){Marino}, {Milone}, {Piotto}, {Villanova},
  {Bedin}, {Bellini}, \& {Renzini}}]{marino2009}
{Marino}, A.~F., {Milone}, A.~P., {Piotto}, G., {et~al.} 2009, \aap, 505, 1099

\bibitem[{{Marino} {et~al.}(2014){Marino}, {Milone}, {Przybilla}, {Bergemann},
  {Lind}, {Asplund}, {Cassisi}, {Catelan}, {Casagrande}, {Valcarce}, {Bedin},
  {Cort{\'e}s}, {D'Antona}, {Jerjen}, {Piotto}, {Schlesinger}, {Zoccali}, \&
  {Angeloni}}]{marino_2014}
{Marino}, A.~F., {Milone}, A.~P., {Przybilla}, N., {et~al.} 2014, \mnras, 437,
  1609

\bibitem[{{Marino} {et~al.}(2019{\natexlab{a}}){Marino}, {Milone}, {Renzini},
  {D'Antona}, {Anderson}, {Bedin}, {Bellini}, {Cordoni}, {Lagioia}, {Piotto},
  \& {Tailo}}]{marino_2019}
{Marino}, A.~F., {Milone}, A.~P., {Renzini}, A., {et~al.} 2019{\natexlab{a}},
  \mnras, 487, 3815

\bibitem[{{Marino} {et~al.}(2019{\natexlab{b}}){Marino}, {Milone}, {Sills},
  {Yong}, {Renzini}, {Bedin}, {Cordoni}, {D'Antona}, {Jerjen}, {Karakas},
  {Lagioia}, {Piotto}, \& {Tailo}}]{Marino2019ApJ...887...91M}
{Marino}, A.~F., {Milone}, A.~P., {Sills}, A., {et~al.} 2019{\natexlab{b}},
  \apj, 887, 91

\bibitem[{{Marino} {et~al.}(2012){Marino}, {Milone}, {Sneden}, {Bergemann},
  {Kraft}, {Wallerstein}, {Cassisi}, {Aparicio}, {Asplund}, {Bedin}, {Hilker},
  {Lind}, {Momany}, {Piotto}, {Roederer}, {Stetson}, \&
  {Zoccali}}]{marino2012m22}
{Marino}, A.~F., {Milone}, A.~P., {Sneden}, C., {et~al.} 2012, \aap, 541, A15

\bibitem[{{Marino} {et~al.}(2017){Marino}, {Milone}, {Yong}, {Da Costa},
  {Asplund}, {Bedin}, {Jerjen}, {Nardiello}, {Piotto}, {Renzini}, \&
  {Shetrone}}]{marino_2017}
{Marino}, A.~F., {Milone}, A.~P., {Yong}, D., {et~al.} 2017, \apj, 843, 66

\bibitem[{{Mazzitelli} {et~al.}(1999){Mazzitelli}, {D'Antona}, \&
  {Ventura}}]{mazzitelli_1999}
{Mazzitelli}, I., {D'Antona}, F., \& {Ventura}, P. 1999, \aap, 348, 846

\bibitem[{{Milone} \& {Marino}(2022)}]{Milone_2022}
{Milone}, A.~P. \& {Marino}, A.~F. 2022, Universe, 8, 359

\bibitem[{{Milone} {et~al.}(2019){Milone}, {Marino}, {Bedin}, {Anderson},
  {Apai}, {Bellini}, {Dieball}, {Salaris}, {Libralato}, {Nardiello},
  {Bergeron}, {Burgasser}, {Rees}, {Rich}, \& {Richer}}]{Milone_2019}
{Milone}, A.~P., {Marino}, A.~F., {Bedin}, L.~R., {et~al.} 2019, \mnras, 484,
  4046

\bibitem[{{Milone} {et~al.}(2014){Milone}, {Marino}, {Dotter}, {Norris},
  {Jerjen}, {Piotto}, {Cassisi}, {Bedin}, {Recio Blanco}, {Sarajedini},
  {Asplund}, {Monelli}, \& {Aparicio}}]{milone_2014}
{Milone}, A.~P., {Marino}, A.~F., {Dotter}, A., {et~al.} 2014, \apj, 785, 21

\bibitem[{{Milone} {et~al.}(2013){Milone}, {Marino}, {Piotto}, {Bedin},
  {Anderson}, {Aparicio}, {Bellini}, {Cassisi}, {D'Antona}, {Grundahl},
  {Monelli}, \& {Yong}}]{milone_2013}
{Milone}, A.~P., {Marino}, A.~F., {Piotto}, G., {et~al.} 2013, \apj, 767, 120

\bibitem[{{Milone} {et~al.}(2015){Milone}, {Marino}, {Piotto}, {Renzini},
  {Bedin}, {Anderson}, {Cassisi}, {D'Antona}, {Bellini}, {Jerjen},
  {Pietrinferni}, \& {Ventura}}]{milone_2015}
{Milone}, A.~P., {Marino}, A.~F., {Piotto}, G., {et~al.} 2015, \apj, 808, 51

\bibitem[{{Milone} {et~al.}(2018){Milone}, {Marino}, {Renzini}, {D'Antona},
  {Anderson}, {Barbuy}, {Bedin}, {Bellini}, {Brown}, {Cassisi}, {Cordoni},
  {Lagioia}, {Nardiello}, {Ortolani}, {Piotto}, {Sarajedini}, {Tailo}, {van der
  Marel}, \& {Vesperini}}]{milone_2018}
{Milone}, A.~P., {Marino}, A.~F., {Renzini}, A., {et~al.} 2018, \mnras, 481,
  5098

\bibitem[{{Milone} {et~al.}(2012{\natexlab{a}}){Milone}, {Piotto}, {Bedin},
  {Aparicio}, {Anderson}, {Sarajedini}, {Marino}, {Moretti}, {Davies},
  {Chaboyer}, {Dotter}, {Hempel}, {Mar{\'{\i}}n-Franch}, {Majewski}, {Paust},
  {Reid}, {Rosenberg}, \& {Siegel}}]{milone_2012c}
{Milone}, A.~P., {Piotto}, G., {Bedin}, L.~R., {et~al.} 2012{\natexlab{a}},
  \aap, 540, A16

\bibitem[{{Milone} {et~al.}(2012{\natexlab{b}}){Milone}, {Piotto}, {Bedin},
  {Cassisi}, {Anderson}, {Marino}, {Pietrinferni}, \& {Aparicio}}]{milone_2012}
{Milone}, A.~P., {Piotto}, G., {Bedin}, L.~R., {et~al.} 2012{\natexlab{b}},
  \aap, 537, A77

\bibitem[{{Milone} {et~al.}(2012{\natexlab{c}}){Milone}, {Piotto}, {Bedin},
  {King}, {Anderson}, {Marino}, {Bellini}, {Gratton}, {Renzini}, {Stetson},
  {Cassisi}, {Aparicio}, {Bragaglia}, {Carretta}, {D'Antona}, {Di Criscienzo},
  {Lucatello}, {Monelli}, \& {Pietrinferni}}]{milone_2012b}
{Milone}, A.~P., {Piotto}, G., {Bedin}, L.~R., {et~al.} 2012{\natexlab{c}},
  \apj, 744, 58

\bibitem[{{Milone} {et~al.}(2017){Milone}, {Piotto}, {Renzini}, {Marino},
  {Bedin}, {Vesperini}, {D'Antona}, {Nardiello}, {Anderson}, {King}, {Yong},
  {Bellini}, {Aparicio}, {Barbuy}, {Brown}, {Cassisi}, {Ortolani}, {Salaris},
  {Sarajedini}, \& {van der Marel}}]{milone_2017}
{Milone}, A.~P., {Piotto}, G., {Renzini}, A., {et~al.} 2017, \mnras, 464, 3636

\bibitem[{{Moehler} {et~al.}(2004{\natexlab{a}}){Moehler}, {Sweigart},
  {Landsman}, {Hammer}, \& {Dreizler}}]{moehler2004}
{Moehler}, S., {Sweigart}, A.~V., {Landsman}, W.~B., {Hammer}, N.~J., \&
  {Dreizler}, S. 2004{\natexlab{a}}, \aap, 415, 313

\bibitem[{{Moehler} {et~al.}(2004{\natexlab{b}}){Moehler}, {Sweigart},
  {Landsman}, {Hammer}, \& {Dreizler}}]{Moehler2004A&A...415..313M}
{Moehler}, S., {Sweigart}, A.~V., {Landsman}, W.~B., {Hammer}, N.~J., \&
  {Dreizler}, S. 2004{\natexlab{b}}, \aap, 415, 313

\bibitem[{{Momany} {et~al.}(2004){Momany}, {Bedin}, {Cassisi}, {Piotto},
  {Ortolani}, {Recio-Blanco}, {De Angeli}, \& {Castelli}}]{momany_2004}
{Momany}, Y., {Bedin}, L.~R., {Cassisi}, S., {et~al.} 2004, \aap, 420, 605

\bibitem[{{Momany} {et~al.}(2012){Momany}, {Saviane}, {Smette}, {Bayo},
  {Girardi}, {Marconi}, {Milone}, \& {Bressan}}]{momany2012a}
{Momany}, Y., {Saviane}, I., {Smette}, A., {et~al.} 2012, \aap, 537, A2

\bibitem[{{Nardiello} {et~al.}(2018){Nardiello}, {Libralato}, {Piotto},
  {Anderson}, {Bellini}, {Aparicio}, {Bedin}, {Cassisi}, {Granata}, {King},
  {Lucertini}, {Marino}, {Milone}, {Ortolani}, {Platais}, \& {van der
  Marel}}]{nardiello_2018}
{Nardiello}, D., {Libralato}, M., {Piotto}, G., {et~al.} 2018, \mnras, 481,
  3382

\bibitem[{{Piotto} {et~al.}(2007){Piotto}, {Bedin}, {Anderson}, {King},
  {Cassisi}, {Milone}, {Villanova}, {Pietrinferni}, \& {Renzini}}]{piotto_2007}
{Piotto}, G., {Bedin}, L.~R., {Anderson}, J., {et~al.} 2007, \apjl, 661, L53

\bibitem[{{Piotto} {et~al.}(2015){Piotto}, {Milone}, {Bedin}, {Anderson},
  {King}, {Marino}, {Nardiello}, {Aparicio}, {Barbuy}, {Bellini}, {Brown},
  {Cassisi}, {Cool}, {Cunial}, {Dalessandro}, {D'Antona}, {Ferraro}, {Hidalgo},
  {Lanzoni}, {Monelli}, {Ortolani}, {Renzini}, {Salaris}, {Sarajedini}, {van
  der Marel}, {Vesperini}, \& {Zoccali}}]{piotto_2015}
{Piotto}, G., {Milone}, A.~P., {Bedin}, L.~R., {et~al.} 2015, \aj, 149, 91

\bibitem[{{Renzini} {et~al.}(2015){Renzini}, {D'Antona}, {Cassisi}, {King},
  {Milone}, {Ventura}, {Anderson}, {Bedin}, {Bellini}, {Brown}, {Piotto}, {van
  der Marel}, {Barbuy}, {Dalessandro}, {Hidalgo}, {Marino}, {Ortolani},
  {Salaris}, \& {Sarajedini}}]{renzini_2015}
{Renzini}, A., {D'Antona}, F., {Cassisi}, S., {et~al.} 2015, \mnras, 454, 4197

\bibitem[{{Rood}(1970)}]{rood1970}
{Rood}, R.~T. 1970, \apj, 161, 145

\bibitem[{{Salaris} {et~al.}(2008){Salaris}, {Cassisi}, \&
  {Pietrinferni}}]{Salaris2008ApJ...678L..25S}
{Salaris}, M., {Cassisi}, S., \& {Pietrinferni}, A. 2008, \apjl, 678, L25

\bibitem[{{Sandage} \& {Wildey}(1967)}]{sw1967}
{Sandage}, A. \& {Wildey}, R. 1967, \apj, 150, 469

\bibitem[{{Schwarzschild} \& {H{\"a}rm}(1962)}]{sh1962}
{Schwarzschild}, M. \& {H{\"a}rm}, R. 1962, \apj, 136, 158

\bibitem[{{Sweigart}(1997)}]{Sweigart1997fbs..conf....3S}
{Sweigart}, A.~V. 1997, in The Third Conference on Faint Blue Stars, ed.
  A.~G.~D. {Philip}, J.~{Liebert}, R.~{Saffer}, \& D.~S. {Hayes}, 3

\bibitem[{{Tailo} {et~al.}(2017){Tailo}, {D'Antona}, {Milone}, {Bellini},
  {Ventura}, {Di Criscienzo}, {Cassisi}, {Piotto}, {Salaris}, {Brown},
  {Vesperini}, {Bedin}, {Marino}, {Nardiello}, \& {Anderson}}]{tailo_2017}
{Tailo}, M., {D'Antona}, F., {Milone}, A.~P., {et~al.} 2017, \mnras, 465, 1046

\bibitem[{{Tailo} {et~al.}(2015){Tailo}, {D'Antona}, {Vesperini}, {di
  Criscienzo}, {Ventura}, {Milone}, {Bellini}, {Dotter}, {Decressin},
  {D'Ercole}, {Caloi}, \& {Capuzzo-Dolcetta}}]{tailo_2015}
{Tailo}, M., {D'Antona}, F., {Vesperini}, E., {et~al.} 2015, \nat, 523, 318

\bibitem[{{Tailo} {et~al.}(2021){Tailo}, {Milone}, {Lagioia}, {D'Antona},
  {Jang}, {Vesperini}, {Marino}, {Ventura}, {Caloi}, {Carlos}, {Cordoni},
  {Dondoglio}, {Mohandasan}, {Nastasio}, \& {Legnardi}}]{tailo_2021}
{Tailo}, M., {Milone}, A.~P., {Lagioia}, E.~P., {et~al.} 2021, \mnras, 503, 694

\bibitem[{{Tailo} {et~al.}(2020){Tailo}, {Milone}, {Lagioia}, {D'Antona},
  {Marino}, {Vesperini}, {Caloi}, {Ventura}, {Dondoglio}, \&
  {Cordoni}}]{tailo_2020}
{Tailo}, M., {Milone}, A.~P., {Lagioia}, E.~P., {et~al.} 2020, \mnras, 498,
  5745

\bibitem[{{van den Bergh}(1967)}]{vdb1967}
{van den Bergh}, S. 1967, \aj, 72, 70

\bibitem[{{VandenBerg} {et~al.}(2013){VandenBerg}, {Brogaard}, {Leaman}, \&
  {Casagrand e}}]{VandenBerg_2013}
{VandenBerg}, D.~A., {Brogaard}, K., {Leaman}, R., \& {Casagrand e}, L. 2013,
  \apj, 775, 134

\bibitem[{{Ventura} {et~al.}(1998){Ventura}, {Zeppieri}, {Mazzitelli}, \&
  {D'Antona}}]{ventura_1998}
{Ventura}, P., {Zeppieri}, A., {Mazzitelli}, I., \& {D'Antona}, F. 1998, \aap,
  334, 953

\bibitem[{{Villanova} {et~al.}(2010){Villanova}, {Geisler}, \&
  {Piotto}}]{villanova2010}
{Villanova}, S., {Geisler}, D., \& {Piotto}, G. 2010, \apjl, 722, L18

\bibitem[{{Yong} {et~al.}(2008){Yong}, {Grundahl}, {Johnson}, \&
  {Asplund}}]{Yong2008ApJ...684.1159Y}
{Yong}, D., {Grundahl}, F., {Johnson}, J.~A., \& {Asplund}, M. 2008, \apj, 684,
  1159

\bibitem[{{Yong} {et~al.}(2003){Yong}, {Grundahl}, {Lambert}, {Nissen}, \&
  {Shetrone}}]{Yong2003A&A...402..985Y}
{Yong}, D., {Grundahl}, F., {Lambert}, D.~L., {Nissen}, P.~E., \& {Shetrone},
  M.~D. 2003, \aap, 402, 985

\bibitem[{{Yong} {et~al.}(2005){Yong}, {Grundahl}, {Nissen}, {Jensen}, \&
  {Lambert}}]{Yong2005}
{Yong}, D., {Grundahl}, F., {Nissen}, P.~E., {Jensen}, H.~R., \& {Lambert},
  D.~L. 2005, \aap, 438, 875

\end{thebibliography}
%

\end{document}